\documentclass[aoas,MSNbibl,nameyear,rotating,dvips]{arximspdf}
\usepackage{graphicx}

\doi{10.1214/11-AOAS498}
\volume{6}
\issue{1}
\pubyear{2012}
\firstpage{304}
\lastpage{333}

\makeatletter

\newcommand{\ga}{\operatorname{Ga}}
\newcommand{\po}{\operatorname{Po}}
\newcommand{\Var}{\operatorname{Var}}
\newcommand{\E}{\mathbb{E}}

\makeatother

\begin{document}
\begin{frontmatter}

\title{Estimation and extrapolation of time trends in~registry
data---Borrowing strength from~related~populations}
\runtitle{Correlated multivariate APC models}

\begin{aug}
\author[A]{\fnms{Andrea} \snm{Riebler}\corref{}\thanksref{t1,t2}\ead[label=e1]{andrea.riebler@ifspm.uzh.ch}},
\author[A]{\fnms{Leonhard} \snm{Held}\thanksref{t1}\ead[label=e2]{leonhard.held@ifspm.uzh.ch}}
\and
\author[B]{\fnms{H{\aa}vard} \snm{Rue}\ead[label=e3]{havard.rue@math.ntnu.no}}
\runauthor{A. Riebler, L. Held and H. Rue}
\affiliation{University of Zurich, University of Zurich
and~Norwegian~University~of~Science~and~Technology}
\address[A]{A. Riebler\\
L. Held\\
Division of Biostatistics\\
ISPM, University of Zurich\\
Hirschengraben 84\\
CH-8001 Zurich\\
Switzerland\\
\printead{e1}\\
\phantom{E-mail: }\printead*{e2}}
\address[B]{H. Rue\\
Department of Mathematical Sciences\\
Norwegian University of Science\\
\quad and Technology\\
N-7491 Trondheim\\
Norway\\
\printead{e3}}
\end{aug}

\thankstext{t1}{Supported by the
Swiss National Science Foundation.}

\thankstext{t2}{Supported by the Research Council of Norway.}

\received{\smonth{7} \syear{2010}}
\revised{\smonth{5} \syear{2011}}

%
\begin{abstract}
To analyze and project age-specific mortality or morbidity rates
age-period-cohort (APC) models are very popular. Bayesian
approa\-ches facilitate estimation and improve predictions by
assigning smoo\-thing priors to age, period and cohort effects.
Adjustments for overdispersion are straightforward using additional
random effects. When rates are further stratified, for example, by
countries, multivariate APC models can be used, where differences of
stratum-specific effects are interpretable as log relative risks. Here,
we incorporate correlated stratum-specific smoothing priors and
correlated overdispersion parameters into the multivariate APC model,
and use Markov chain Monte Carlo and integrated nested Laplace
approximations for inference. Compared to a model without correlation,
the new approach may lead to more precise relative risk estimates, as
shown in an application to chronic obstructive pulmonary disease
mortality in three regions of England and Wales. Furthermore, the
imputation of missing data for one particular stratum may be improved,
since the new approach takes advantage of the remaining strata if the
corresponding observations are available there. This is shown in an
application to female mortality in Denmark, Sweden and Norway from the
20th century, where we treat for each country in turn either the first
or second half of the observations as missing and then impute the
omitted data. The projections are compared to those obtained from a
univariate APC model and an extended Lee--Carter demographic forecasting
approach using the proper Dawid--Sebastiani scoring rule.
\end{abstract}

%

\end{frontmatter}

\section{Introduction}
Most developed countries have national health registers to routinely collect
demographic rates. Age-period-cohort (APC) models are commonly
used to analyze and project mortality or morbidity rates,
in which effects related to the age of an individual, calendar
time (period) and the generation (cohort) can reasonably be assumed to
be present.
When several of such register data sets are available, for example, for
different countries,
each data set could be analyzed separately by a univariate APC model.
However, for comparable
strata, similar unobservable factors are likely to act on
the different time dimensions (age, period, cohort), so that a
multivariate APC analysis
may seem more appropriate [\citet{hansell-etal-2003},
\citet{hansell-phd-2004}, \citet{jacobsen-etal-2004},
\citet{riebler-held-2010}].

A quirk of APC models is the obvious linear dependence of age, period
and cohort
effects leading to a well-known identifiability problem. Over the
last decades several proposals, ranging from the
specification of additional identifying restrictions to the definition
of estimable functions, have been
made to solve the identifiability problem; see, for example,
\citet{fienberg-mason-1979},
\citet{osmond-gardner-1982}, \citet{holford-1983},
\citet{robertson-boyle-1986},
\citet{clayton-schifflers-1987},
\citet{holford-1992}, \citet{fu-2000}, \citet{yang-etal-2004}
or \citet{kuang-etal-2008}.
Provided that at least one set of age,
period or cohort effects is forced to be identical across strata (which
is often a
plausible assumption), differences of stratum-specific effects in the
multivariate APC model
are identifiable without further identifying restrictions.
They can be interpreted as log relative risks, so that
heterogeneous time trends, for example, across gender
[\citet{riebler-etal-2011}] or geographical
regions [\citet{hansell-phd-2004}, \citet
{riebler-held-2010}], can be analyzed.

Bayesian APC analyses have become very popular in the last years; see,
for example,
\citet{nakamura-1986}, \citet{berzuini-clayton-1994},
\citet{besag-etal-1995}, \citet{ogata-etal-2000},
\citet{knorrHeld-rainer-2001}, \citet{bray-etal-2001},
\citet{bray-2002}, \citet{baker-bray-2005},
\citet{schmid-held-2007}, \citet{riebler-held-2010}.
As effects adjacent in time are likely to be similar, smoothing priors
are typically assumed
for age, period and cohort effects.
\citet{nakamura-1986} used first-order autoregressive priors, while
\citet{berzuini-clayton-1994} and \citet{besag-etal-1995}
proposed to use
second-order random walks.
The second-order random walk is a discrete-time analogue of a cubic
smoothing spline [\citet{Fahrmeir-tutz-2001}].
This prior is defined on the identifiable second differences, a well-known
measure of curvature, and penalizes deviations from a linear trend
[\citet{fienberg-mason-1979}, \citet{clayton-schifflers-1987}].
The degree of smoothness is controlled by an unknown smoothing parameter.
Using smoothing priors, overfitting cohorts, which by design are
sparsely represented, is
avoided [\citet{besag-etal-1995}].

When age group and period intervals are of different length, an
additional identifiability
problem may induce artificial cyclical patterns in the parameter
estimates of
uni- and multivariate APC models; see \citet{holford-2006} and
\citet
{riebler-held-2010}, respectively.
However, this problem can be solved by applying smoothing functions,
such as second-order random walks or penalized splines
[\citet{holford-2006}].
The assumed smoothness of age, period and cohort effects
can also be exploited for providing projections, as the effects can
easily be extrapolated into both
the future and past [\citet{knorrHeld-rainer-2001}, \citet
{bray-2002}].
In a hierarchical Bayesian model, additional random effects can be included
to account for heterogeneity without temporal structure
[\citet{berzuini-clayton-1994}, \citet{besag-etal-1995}].

\citet{riebler-held-2010} assumed independent smoothing priors for
stra\-tum-specific effects in multivariate APC models. Here, we propose
to link
stratum-specific smoothing priors. The new approach leads to a
multivariate correlated random walk, where the
joint precision matrix is defined as the Kronecker product of
the inverse of a uniform correlation matrix and the precision matrix of
the univariate second random walk.
Inference is done using Markov chain Monte Carlo (MCMC) and integrated
nested Laplace approximations
[\citet{rue-etal-2009}].

The new specification can be regarded as shrinking the stra\-tum-specific
parameters toward some common trend.
Indeed, an alternative model formulation would introduce a common
period effect, say, modeled via a second
order random walk and additionally
independent second order random walks for each stratum. While this
formulation has
two variance parameters, it in fact induces correlation between the
stratum-specific increments, which
are defined as the sum of the common innovation and the
stratum-specific innovations,
so that it can be translated into a multivariate random walk with one
variance and one correlation
parameter.

In time series analysis, the use
of multivariate random walks plays a fundamental role in multivariate
modeling [\citet{Harvey-1990}].
The multivariate random walk is an example of an intrinsic multivariate
Gaussian Markov random
field (GMRF) model [\citet{GMRFbook}]. Multivariate GMRF models with
conditional autoregressive (CAR) structure
are sometimes called multivariate CAR (MCAR) models; see, for example,
\citet{gelfand-vounatsou-2003} or \citet
{carlin-banerjee-2003}. Proper
multivariate GMRF models
have been introduced by \citet{mardia-1988}.
\citet{greco-trivisano-2009} applied MCAR models to handle
general forms
of spatial dependence
occurring in multivariate spatial modeling of area data. \citet
{lagazio-etal-2003} and \citet{schmid-held-2004}
used Kronecker product precision matrices to model different types of
space--time interactions in
spatial APC models [\citet{knorrHeld-2000}].
However, as far as we know, correlated second order random walks have
never been used in
multivariate APC models.

We further propose the incorporation of correlated overdispersion parameters
to model unobserved risk factors without temporal structure but acting
simultaneously on the
different strata. The use of correlated overdispersion parameters is
similar in spirit to
seemingly unrelated regressions, where single
regression equations are linked by correlated error terms
[\citet{Harvey-1990}].

Through the introduction of correlation in the prior distribution
the effective degrees of freedom are reduced whenever similar behavior
in the different
strata exists. Hence, the precision of relative risks may be improved.
Furthermore, the approach is useful to predict missing records in one
particular stratum
if the corresponding data are available for the remaining strata.
This might be the case for historical data if the collection of
demographic rates started not at the same time in
different strata.
Consider, for example, Switzerland, where each canton (administrative
unit) is separately responsible for the implementation of health-policy
instruments,
so that cancer is registered on a cantonal level
[\citet{ess-etal-2010}, \citet{krebs}].
The first Swiss cancer registration system started in 1970 in the
canton of
Geneva followed by registers in the cantons of Vaud
and Neuch{\^a}tel in 1974 [\citet{krebs}]. Compared to other cantons
without explicit cancer
registration, extensive cancer analyses have been performed for these cantons;
see, for example, Levi et~al. (\citeyear
{levi-etal-1993,levi-etal-1998,levi-etal-2002}),
\citet{verkooijhen-etal-2003}. Today most cantons have cancer
registers and it
is planned
that within the next years the entire Swiss population will be captured by a cancer
registration system [\citet{krebs}].
Our method can be used to impute missing data for cantons with a~younger cancer registration
system taking advantage of other cantons with a~longer collection
period. Thus, important
insight into cancer progression for all cantons could be gained.
A different aspect might be varying collection intervals in different
regions, where in some regions data are collected on a yearly basis,
say, and in other regions on a
five-year basis. Here, the correlated multivariate APC approach may be
used to impute the
rates for the missing years.
In this paper, we demonstrate the ability to impute missing data units
in a cross-prediction study of
female mortality in Scandinavia.

The paper is organized as follows. Section \ref{secapp} introduces the
two applications
presented in this paper. In Section \ref{seccmapc} we review
multivariate APC models and
introduce our extended correlated approach (Section \ref{secinference}).
Then we present details on the implementation (Section \ref
{secimplementation}).
In Section \ref{secapplications} we present the results of the two
applications.
Our findings are summarized in Section \ref{secdiscussion}.

\section{Applications}\label{secapp}
\subsection{Analysis of heterogeneous time trends in COPD mortality
among males in England and Wales}\label{seccopd}
We reanalyze male mortality data on chronic obstructive pulmonary
disease (COPD) in three regions
of England and Wales: Greater London, conurbations excluding Greater
London and rural areas (nonconurbations).
COPD is one of the most common lung diseases making it hard
to breathe as a consequence of limited air flow. One of the main causes
of COPD is smoking, but also air
pollution, smog, dust~and~che\-mical fumes are relevant risk factors.
While smoking exerts mainly
long-term effects with a lag period of about 20--30 years [\citet
{kazerouni-etal-2004}],
air pollution can cause both long-term (period or cohort)
effects and short-term (period) effects
[\citet{sunyer-2001}, \citet{dockery-pope-1994}].
We focus on short-term effects and the relation between marked air
pollution events
and changes in COPD mortality. For all regions data are available on an
annual basis from 1950--1999 for seven
age groups: 15--24, 25--34$, \ldots, 75+$ [\citet{hansell-etal-2003},
\citet{hansell-phd-2004}].
\citet{riebler-held-2010} analyzed heterogeneous time trends in these
data using an
uncorrelated multivariate APC model with common age effects. We will
compare their results with those
obtained from a model with correlated stratum-specific period, cohort
and overdispersion parameters.

\subsection{Extrapolation of overall mortality of Scandinavian females}
\begin{figure}

\includegraphics{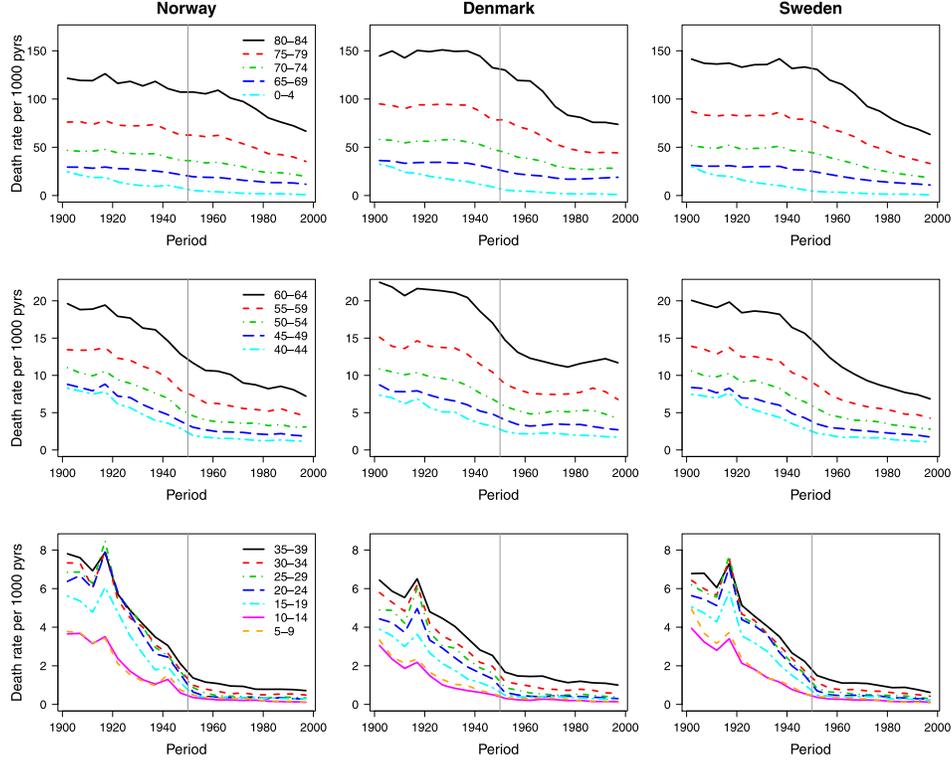}

\caption{Female death rates per $1\mbox{,}000$ person-years (pyrs) in Norway,
Denmark and Sweden by
age from 1900 to 1999. The vertical line divides the time into equally
sized parts.}
\label{figscan}
\end{figure}

All data were obtained from the \citet{hmd-2010}. The number of deaths
are stratified by 5-year groups,
for all Norwegian, Danish and Swedish women aged 0--84 in the period 1900--1999,
leading to 17 age groups (0--4, 5--$9, \ldots, 80$--84) and 20
periods (1900--$1904,\ldots, 1995$--1999).
Figure \ref{figscan} shows the death rates per $1\mbox{,}000$
person-years for all three
countries stratified by 5-year age groups. To obtain person-years, we
used the yearly
population sizes available for the same age groups
and based on the 1st of January. We used linear interpolation to get
mid-year estimates and then
added up the resulting quantities to obtain person-years for
1900--$1904, \ldots, 1995$--1999.
%
%
The rates of all three countries show a very similar progression. The
peak in mortality
in the 1915--1919 period, present, in particular, among young adults,
is supposed to be related to the 1918--1919 Spanish
flu pandemic which killed about 50 million people worldwide, with most
deaths occurring among young adults [\citet{andreasen-etal-2008}].
During the summer of 1918 there
were strong influenza waves in Denmark, Sweden and Norway
[\citet{andreasen-etal-2008}, \citet{kolte-etal-2008}].

We divide the calendar period into two equally sized parts (see
Figure~\ref{figscan}).
For either the first or second half of the 20th century, all
observations from one particular country
are treated as missing. The omitted data are then predicted exploiting
the information
provided by the complete data sets of the other two countries. This procedure
is repeated for all countries and hence termed cross-prediction.
Thus, we cannot only assess the ability to project particular events
present for all countries, such as
the Spanish-flu pattern, but also analyze the prediction quality for
years without such specific events.
The probabilistic projections are compared to those obtained from a
univariate APC model
and to those from an extended Lee--Carter demographic forecasting model
[\citet{lee-carter-1992}].

The data and code used in this application are provided in the
\hyperref[suppA]{Supplemen-} \hyperref[suppA]{tary Material}
[\citet{riebler-etal-2011supp}].

\section{The correlated multivariate APC model}\label{seccmapc}
Let $y_{ijr}$ denote the number of deaths observed for age group
$i$ $(i = 1, \ldots, I)$, period $j$ ($j=1, \ldots, J$) and stratum $r$
($r=1, \ldots, R$).
In both of our applications $r$ represents a~geographical
region (either a region in England and Wales or a Scandinavian country).
Deaths can be regarded as events arising from a Poisson process.
Hence, $y_{ijr}$ can be interpreted as the number of events that have
occurred during an
exposure period of $n_{ijr}$ person-years, in which the occurrence rate
is assumed to be
$\lambda_{ijr}$ per person-year. Thus, $y_{ijr}$ is Poisson
distributed with
rate $n_{ijr}\lambda_{ijr}$, where $n_{ijr}$ is known
[\citet{armitage-1966}, \citet{brillinger-1986}].
In the most general formulation of the multivariate APC model, the
linear predictor is
\[
\eta_{ijr} = \log(\lambda_{ijr}) = \mu_r + \theta_{ir} + \varphi_{jr}
+ \psi_{kr}.
\]
Here, $\mu_{r}$ is the stratum-specific intercept, and
$\theta_{ir}$, $\varphi_{jr}$ and $\psi_{kr}$ are stratum-specific age,
period and cohort effects, respectively. The cohort index $k$ is a~linear function of
the age index $i$ and the period index $j$. If the time interval
widths
of age group
and period are equal, then $k=(I-i)+j$. If age group intervals are $M$
times wider
than period intervals, as is the case in the first application
(Section \ref{seccopd}) where $M=10$, then $k = M \times(I-i)+j$
[\citet{heuer-1997}].
We apply the usual constraints,
\mbox{$\sum_{i=1}^I\theta_{ir} = \sum_{j=1}^J \varphi_{jr} = \sum
_{k=1}^K \psi
_{kr} = 0$}
for $r=1, \ldots, R$, to ensure identifiability of the stratum-specific
intercepts. However, parameter estimates are still not identifiable
without imposing
additional constraints [\citet{fienberg-mason-1979},
\citet{holford-1983}]. In
contrast, second differences
of parameter estimates, for example, $\theta_{ir} - 2\theta_{{i-1}r} +
\theta_{{i-2}r}$,
are not affected by the identifiability problem
and can be uniquely determined [\citet{fienberg-mason-1979},
\citet{clayton-schifflers-1987}].
Furthermore, stratum-specific differences, for example, $\theta_{ir_1}
- \theta_{ir_2}$ with $r_1 \ne r_2$,
are identifiable (absent an additional constraint),
provided that at least one of the three time effects (age, period,
cohort) is common across strata
[\citet{riebler-held-2010}].

\subsection{Bayesian inference} \label{secinference}

In a Bayesian context, we work with a hierarchical model in which
prior distributions need to be assigned to all parameters.
We use independent flat priors for each stratum-specific
intercept $\mu_{r}$.
\citet{riebler-held-2010} assigned independent smoothing priors to
the age effects $\bolds{\theta} = (\theta_1, \ldots, \theta_I)^\top$,
each stratum-specific set of period effects $\bolds{\varphi}_r =
(\varphi
_{1r}, \ldots, \varphi_{Jr})^\top$
and cohort effects $\bolds{\psi}_r = (\psi_{1r}, \ldots, \psi
_{Kr})^\top$,
$r=1, \ldots, R$, in
a model with common age effects.
Consider, for example, the period effects for a specific stratum $r$.
The random
walk of second order (RW2) is a~smoothing prior based on second differences and penalizes deviations
from a linear trend.
This improper prior can be written as
\begin{eqnarray*}
f(\bolds{\varphi}_r|\kappa_\varphi) &\propto& \kappa_\varphi
^{(J-2)/2}\exp
\Biggl(-\frac{\kappa_\varphi}{2} \sum_{j=3}^J\bigl(\bigl(\varphi_{jr} -\varphi
_{(j-1)r}\bigr) - \bigl(\varphi_{(j-1)r} - \varphi_{(j-2)r}\bigr)\bigr)^2\Biggr)\\
&=& \kappa_\varphi^{(J-2)/2}\exp\biggl(- \frac{1}{2} \bolds{\varphi
}_r^\top\mathbf{P}_\varphi\bolds{\varphi}_r \biggr)
\end{eqnarray*}
with precision matrix $\mathbf{P}_\varphi$, which depends on an unknown
precision
parameter $\kappa_\varphi$:
\[
\mathbf{P}_\varphi= \kappa_\varphi
\pmatrix{
1 & -2 & 1 &&&&\cr
-2& 5 & -4 & 1&&&\cr
1& -4 & 6& -4 &1 &&\cr
&\ddots&\ddots&\ddots&\ddots&\ddots&\cr
&&1& -4& 6& -4 &1\cr
&&&1& -4 & 5 & -2\cr
&&&&1&-2&1}.
\]

Here, we propose the use of correlated smoothing priors for stratum-specific
time effects.
Let $\mathbf{C} = (1-\rho) \mathbf{I} + \rho\mathbf{J}$ denote
a uniform correlation matrix, where $\rho$ is the unknown correlation
parameter,
$\mathbf{I}$ the identity matrix and~$\mathbf{J}$ a matrix of
ones.
The random walks of the stratum-specific period effects
$\bolds{\varphi}_1, \ldots, \bolds{\varphi}_R$ can be correlated
using the stacked vector $\tilde{\bolds{\varphi}} = (\bolds{\varphi
}_1^\top,
\ldots, \bolds{\varphi}_R^\top)^\top$:
\begin{eqnarray*}
f(\tilde{\bolds{\varphi}}|\mathbf{C}_\varphi, \kappa_\varphi)
&\propto&
(|\mathbf{C}^{-1}_\varphi\otimes\mathbf{P}_\varphi|^\star)^
{1/2} \exp\bigl(-\tfrac{1}{2}
\tilde{\bolds{\varphi}}^\top\{\mathbf{C}^{-1}_\varphi\otimes\mathbf
{P}_\varphi\}\tilde{\bolds{\varphi}}\bigr)\\
&=& |\mathbf{C}^{-1}_\varphi|^{({J-2})/{2}} \cdot(|\mathbf
{P}_\varphi
|^\star)^{R/2} \exp\bigl(-\tfrac{1}{2}
\tilde{\bolds{\varphi}}^\top\{\mathbf{C}^{-1}_\varphi\otimes\mathbf
{P}_\varphi\}\tilde{\bolds{\varphi}}\bigr),
\end{eqnarray*}
where $\otimes$ denotes the Kronecker product and \mbox{$|\cdot|^\star$} the
generalized
determinant defined as the product of all nonzero eigenvalues. The
determinant of~$\mathbf{C}^{-1}_\varphi$ is $[(1+(R-1)\rho_\varphi)(1-\rho
_\varphi
)^{R-1}]^{-1}$; see the proof in
Appendix \ref{secappendix}.
This formulation corresponds to a multivariate RW2 with
correlated increments and is an example for an improper (intrinsic) correlated
GMRF [\citet{gelfand-vounatsou-2003}, \citet{GMRFbook}].

To adjust for heterogeneity, which has no temporal structure but is likely
to exist in the underlying rates, we introduce
stratum-specific latent random effects
$z_{ijr}$ into the linear predictor.
These overdispersion parameters are typically assumed to be independent
Gaussian variables with mean zero and unknown variance $\kappa_z^{-1}$,
that is, $z_{ijr}\stackrel{\mathrm{i.i.d.}}{\sim}\mathcal{N}(0, \kappa_z^{-1})$
[\citet{berzuini-clayton-1994},
\citet{besag-etal-1995}].\vspace*{1pt}
However, when
interpreting these latent effects as unobserved covariates,
it may be plausible that they act partly simultaneously on the
different strata.
Hence, we propose correlated overdispersion parameters and set
$\mathbf{z}_{ij} = (z_{ij1}, \ldots, z_{ijR})^\top\sim
\mathcal{N}(0, \kappa_z^{-1}\mathbf{C}_z)$ for all $i$ and $j$.

All of the up to eight hyperparameters (four precisions and up to four
correlations) are treated as unknown. Suitable gamma-hyperpriors are assigned
to the precisions. As in \citet{knorrHeld-rainer-2001}, we use
$\ga(1, 0.00005)$
for the precisions of age, period and cohort effects and $\ga(1, 0.005)$
for the precision of the overdispersion.

Correlation parameters $\rho$ are reparameterized using the general
Fisher's $z$-transformation
[\citet{Fisher-1958}, page 219]:
%
%
\begin{equation}\label{eqgenFisher}
\rho= \frac{\exp(\rho^\star) -1}{\exp(\rho^\star) + R -1},\qquad \rho
^\star
= \log\biggl(\frac{1+\rho\cdot(R-1)}{1-\rho}\biggr),
\end{equation}
where $\rho^\star$ can take any real value.
It is worth noting that this transformation ensures that $\rho$ only
takes values
within the interval $(-1/(R-1),1)$, so that $\mathbf{C}$ is positive
definite without
imposing an additional constraint. Using $R=2$ in (\ref{eqgenFisher}),
we obtain
\[
\rho= \frac{\exp(\rho^\star) -1}{\exp(\rho^\star) + 1},\qquad
\rho^\star
= \log\biggl(\frac{1+\rho}{1-\rho}\biggr),
\]
which is frequently used for constructing confidence intervals for
$\rho
$ [\citet{konishi-1985}].
Fisher's $z$-transformation is a variance stabilizing transformation.
In a Bayesian context this transformation is of particular interest since
the derivative of a variance stabilizing transformation corresponds to
Jeffreys' prior for the original parameter [\citet{Lehmann-1999},
pages 491 and 492].
For example, for $R=2$, Jeffreys' prior is $\pi(\rho) \propto
1/(1-\rho^2)$,
the derivative of $\log(\frac{1+\rho}{1-\rho})$
[\citet{Lindley-1965}, pages 215--220].

\begin{figure}

\includegraphics{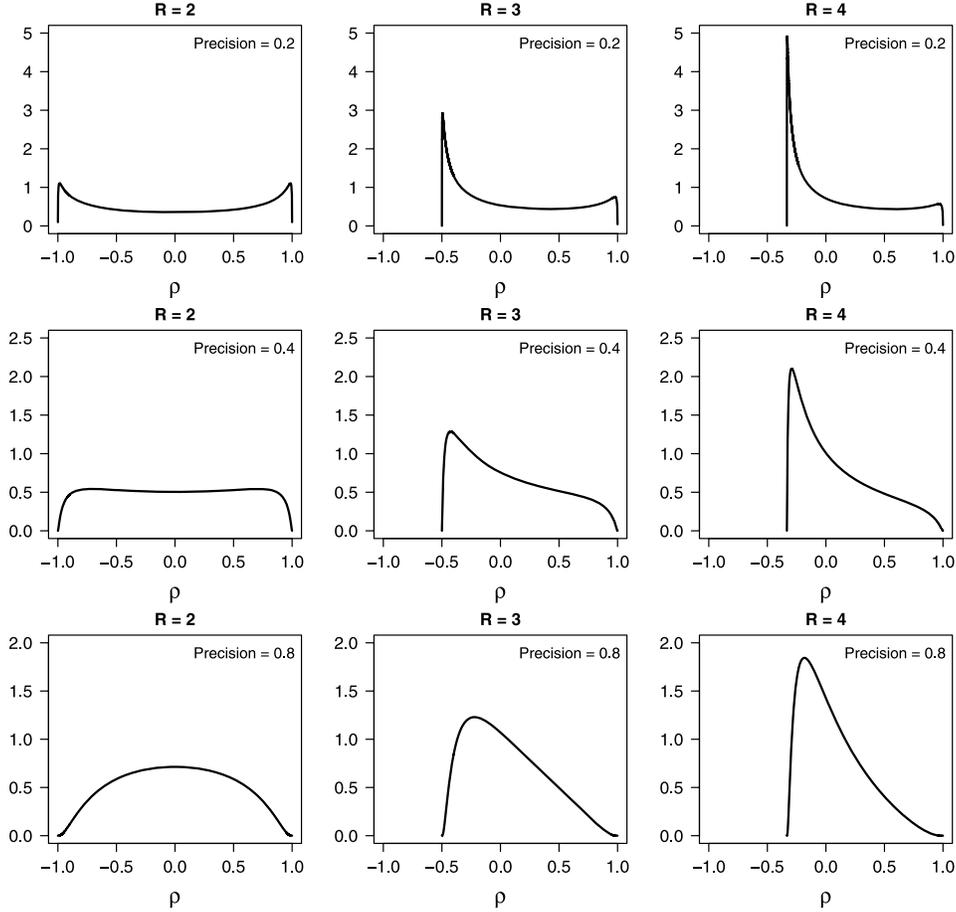}

\caption{Prior distribution for correlation parameter $\rho$ derived
from a zero-mean Gaussian distribution for $\rho^\star$ with three
different values for the precision $\kappa_{\rho^\star}$ (top to
bottom: $0.2$, $0.4$, $0.8$) and three different numbers of strata $R$
(left to right: $R=2$, $R=3$, $R=4$).} \label{figcorPrior}
\end{figure}

We assign a normal prior with mean zero and fixed precision $\kappa
_{\rho^\star}$ to $\rho^\star$.
Thus, the prior probability that $\rho$ is larger than zero is equal to
$0.5$, independent of $R$.
Figure \ref{figcorPrior} shows the resulting prior for $\rho$ for
three different values of
$\kappa_{\rho^\star}$ and three different values of $R$.
%
%
For $R=2$ strata, setting $\kappa_{\rho^\star}$ to $0.2$ corresponds to
a U-shaped prior, $\kappa_{\rho^\star}=0.4$ to a roughly uniform prior
and $\kappa _{\rho ^\star}=0.8$ to a bump-shaped prior for $\rho$;
compare the first column of Figure \ref {figcorPrior}. Note that
$\kappa_{\rho^\star}=0$ corresponds to the improper Jeffreys' prior.
For a larger number of strata, the left boundary for the correlation is
shifted toward zero, resulting in an asymmetric prior distribution
for~$\rho$, since half of the total density is distributed to a smaller
interval, $(-1/(R-1), 0)$. We use $\kappa_{\rho^\star}=0.2$, so that
sufficient probability mass is assigned to the boundary values as well,
making extreme posterior correlation estimates possible.

\subsection{Implementation} \label{secimplementation}
Bayesian inference for the models presented is not straightforward,
since the posterior distribution is not analytically available. The
common tool of choice is MCMC sampling. An alternative is integrated
nested Laplace approximations (INLAs). To compare these two inference
techniques, we implemented correlated multivariate APC models using
both MCMC and INLA. In the first application, we apply INLA and MCMC to
show the almost perfect coincidence of both approaches. Due to the
complexity of the second application resulting in large thinning
intervals and burn-in periods, we only present the results of INLA.

\subsubsection{Analysis with MCMC} \label{secmcmc}
Algorithmic routines based on MCMC are implemented in the low-level
programming language \texttt{C} using the \texttt{GMRFLib} library
[\citet{GMRFbook}]. Following \citet{besag-etal-1995}, we
reparameterize the model from $z_{ijr}$ to $\eta_{ijr}$ to obtain
multivariate normal full conditional distributions for the
stratum-specific intercepts $\bolds{\mu} = (\mu_1, \ldots, \mu_r)^\top$
and all sets of time effects. Block updating allows the proper
incorporation of the sum-to-zero constraints for the time effects. It
is also possible to omit the sum-to-zero constraint for one set of
stratum-specific effects and simultaneously remove the stratum-specific
intercepts $\bolds{\mu}$ from the algorithm. For the precisions Gibbs
sampling is used as well. The vector $\bolds {\eta }_{ij} =
(\eta_{ij1},\ldots, \eta_{ijR})^\top$ has a nonstandard distribution.
It is updated using multivariate Metropolis--Hastings steps with a GMRF
proposal distribution based on a second-order Taylor approximation of
the log likelihood [\citet{GMRFbook}, Section~4.4.1]. For the
correlation parameters Metropolis--Hastings updates based on a random
walk proposal are used, such that acceptance rates around $40\%$ are
achieved. In the application to COPD mortality we use a MCMC run of
$350\mbox {,}000$ iterations, discarding the first $50\mbox{,}000$
iterations and storing every $20$th sample thereafter, resulting in
$15\mbox{,}000$ samples. We have routinely examined convergence and
mixing diagnostics.

\subsubsection{Analysis with INLA} \label{secinla}
\citet{rue-etal-2009} proposed with INLA an alternative
deterministic Bayesian inference approach for latent Gaussian random
field models. INLA replaces time-consuming MCMC sampling with fast and
accurate approximations to the posterior marginal distributions. Some
empirical comparison with MCMC results can be found in
\citet{rue-etal-2009}, \citet{paul-etal-2010} or
\citet{schroedle-etal-2010}. We incorporated correlated GMRF
models into INLA, enabling the analysis of correlated multivariate APC
models based on a uniform correlation structure and using the general
Fisher's $z$-transformation. The methodology is integrated in the
package \texttt{INLA} (see
\href{http://www.r-inla.org/}{www.r-inla.org}) for R
[\citet{r-project}]. For both applications, we use the
\texttt{INLA} package built on 14.03.2011.

\section{Results} \label{secapplications}

\subsection{COPD mortality among males in England and Wales} \label
{secrescopd} 
\begin{figure}

\includegraphics{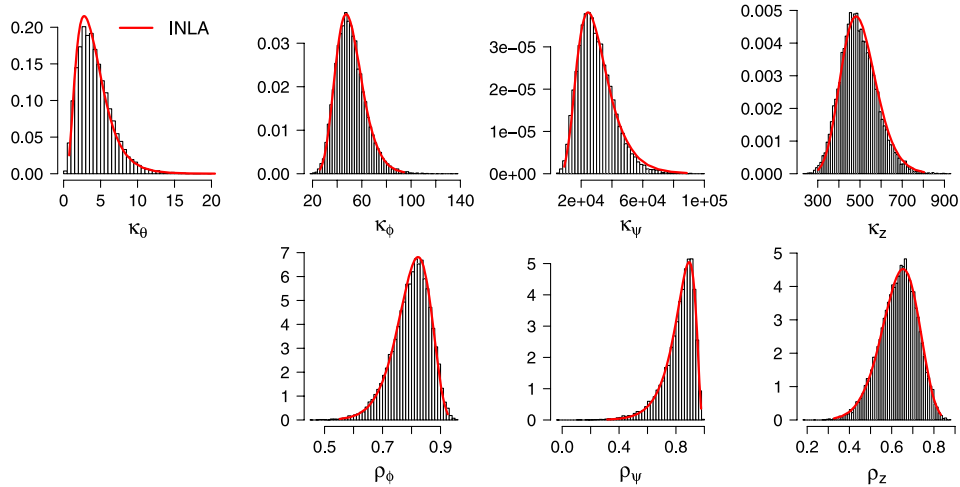}%

\caption{Approximated posterior marginal densities (solid red line) of
precision and correlation parameters
for the multivariate model with joint age effects and correlated
period, cohort and overdispersion parameters
obtained from INLA. Moreover, the corresponding histograms of $15\mbox{,}000$
MCMC samples
obtained from a run with $50\mbox{,}000$ burn-in iterations and a thinning of
$20$ are shown.}
\label{figinlamcmccorApcz}
\end{figure}

We compared the uncorrelated model with joint age-effects,
and region-specific period and cohort effect presented by
\citet{riebler-held-2010} with three different correlated
formulations: (1)~Region-specific period and region-specific cohort
effects are correlated; (2)~the overdispersion parameters are
correlated; (3)~region-specific period effects, region-specific cohort
effects and overdispersion parameters are correlated. To make the
models comparable, we used, in contrast to
\citet{riebler-held-2010}, the same precision for the independent
priors of region-specific period effects and also the same precision
for the independent priors of region-specific cohort effects. For all
models MCMC and INLA produce virtually identical results; see Figure
\ref{figinlamcmccorApcz} for a comparison of precision and correlation
estimates in model 3. The running time of INLA was always less than the
%
%
computation time with MCMC. Inspecting the log marginal likelihood
returned by INLA, the model with correlated period and cohort
effects,
and correlated overdispersion parameters, was classified as the best
model. Despite the improper random walk prior, the log-marginal
likelihood can be used here, as the models are based on the same
underlying latent structure and only differ by the inclusion of
correlation. Furthermore, the correlation estimates $\rho_\varphi,
\rho_\psi$ and $\rho_z$ of model 3 (Figure \ref{figinlamcmccorApcz})
are clearly different from zero, confirming the between-region
dependence.

\begin{figure}

\includegraphics{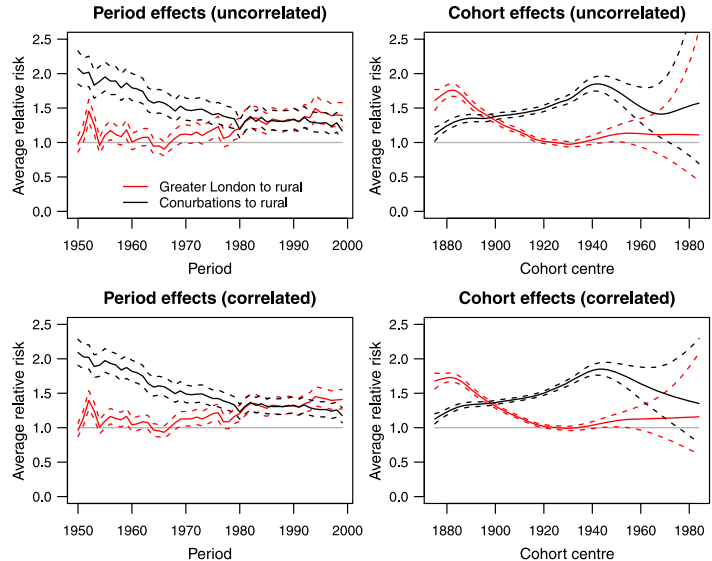}

\caption{Average relative risk of death for Greater London and
conurbations excluding
Greater London compared with nonconurbations, analyzed by a
multivariate model
with joint age effects and no correlations across other parameters
(top), and a multivariate
model with joint age effects and correlated period, cohort and
overdispersion parameters (bottom).
Shown are the median estimates within $95\%$ pointwise credible bands.}
\label{figrrcorVSuncor}
\end{figure}

Figure \ref{figrrcorVSuncor} compares the estimates of average
relative risks obtained
from MCMC for the models with uncorrelated and correlated
region-specific effects and overdispersion parameters, respectively.
The estimates are relative to nonconurbations, where the mortality
rates tend to be the lowest.
The results of both models are very similar. The average relative risk of
period effects shows the typical year-to-year variation, with higher
values in
years of known air pollution events, such as the ``Great Smog'' in
London in 1952.
In the average relative risks of cohort effects, different smoking behavior
may be visible. For a detailed interpretation
of the relative risks we refer to \citet{riebler-held-2010}. Due to
fewer observations,
the credible intervals are getting wider for younger birth cohorts. However,
adjusting for correlation improves the precision of the relative risks
estimates,
in particular, for younger birth cohorts. The average posterior
standard deviation
in the correlated approach is about $20\%$ and $25\%$ smaller for the
average relative risks of the period effects and cohort effects,
respectively.

\subsection{Extrapolation of overall mortality of Scandinavian females}
\label{secscan}
We will first briefly introduce the basic and the extended Lee--Carter
model considered.
Then we will present the results of the predictive model assessment and
compare the
projections obtained by the different approaches.

\subsubsection{The quasi-Poisson version of the Lee--Carter model}
The Lee--Carter model, introduced by \citet{lee-carter-1992} to
forecast mortality in the U.S.,
is one of the best-known methods for mortality forecasting and often
used as a reference
[\citet{booth-2006}, \citet{booth-etal-2006}]. It assumes a~log-bilinear form
\[
\log\frac{y_{ij}}{n_{ij}} = \alpha_i + \beta_i\kappa_j +
\varepsilon_{ij},
\]
where $\alpha_i$ describes the average shape of the age profile,
$\beta
_i$ the age-specific
mortality change from this pattern with time-varying trend $\kappa_j$, and
$\varepsilon_{ij}$ are homoskedastic centered error terms. The
parameters are
constrained to $\sum_j \kappa_j=0$ and $\sum_i \beta_i= 1$.
Forecasting using this model proceeds in two steps: (1) the model
coefficients are estimated; (2)
the time trend $\kappa_j$ is extrapolated based on an ARIMA(0,1,0)
time-series model, that is, a
random walk with drift. This forecasted trend is used to derive the
projected age-specific mortality rates
based on the estimates for $\alpha_i$ and $\beta_i$ from step 1.
Note that only the uncertainty in the time trend $\kappa_j$ is taken
into account in the
projected rates, so that not all variability is captured
[\citet{lee-carter-1992}, \citet{butt-haberman-2009}].

The Lee--Carter model was further developed and embedded in a quasi-Poisson
regression model by \citet{brouhns-etal-2002}. We used the
ilc-package [\citet{butt-haberman-2009}] in R to generate univariate
predictions for the country under consideration based on this extended model.
Since the implementation does not allow to project into the past, we
reversed the
time-scale when predicting data of the first half of the 20th century.

%
%
%

\subsubsection{Predictive model assessment}
$\!\!$Each of the three models (Lee--Carter, univariate APC and multivariate
APC; with the latter two abbreviated as
APC and cMAPC, resp.) generates, for each of the six scenarios
of the cross-prediction procedure, $10 \times17= 170$
probabilistic forecasts for country $r^\star$ under consideration.
We used the mean squared error score to assess the concentration of the
predictive distribution (sharpness).
To assess the statistical consistency between the distributional
forecasts and the observations (calibration),
we calculated prediction intervals at various levels and computed the
empirical coverage probabilities, that is, the proportion of the
prediction intervals
that cover the observed number of cases. To combine sharpness and
calibration in one measure, we further report\vadjust{\goodbreak}
the Dawid--Sebastiani scoring rule (DSS) defined as
\[
\mathrm{DSS} = \biggl(\frac{y_{ijr^\star} - \mu_{ijr^\star}}{\sigma
_{ijr^\star}}\biggr)^2 + 2\log(\sigma_{ijr^\star}),
\]
where $y_{ijr^\star}$ is the observation that realizes, $\mu
_{ijr^\star
}$ the mean and $\sigma_{ijr^\star}$ the standard
deviation of the predictive distribution
[\citet{gneiting-raftery-2007}]. This score has been
proposed as a proper alternative to the predictive
model choice criterion of \citet{gelfand-ghosh-1998} and was also
used by \citet{czado-etal-2009} to assess the predictive
quality of a univariate APC analysis applied to cancer incidence in Germany.
To calculate these quantities, we need to post-process the results
returned by
INLA and the ilc-package.

For the univariate and multivariate APC analysis, INLA returns posterior
summary estimates and posterior marginal densities for the linear predictor
$\eta_{ijr^\star}$, with $i=1, \ldots, 17$ and $j=1, \ldots, 10$ or
$j=11,\ldots,20$, depending
on whether we project the first or last half of the 20th century. The
corresponding estimates for
$\lambda_{ijr^\star}$ are straightforward to derive. For the univariate
Lee--Carter analysis of country $r^\star$, the ilc-package returns
the predicted mortality rate $\lambda_{ijr^\star}$ and
the prediction intervals (symmetric on the log-scale) at a predefined level.

We need the mean $\mu_{ijr^\star}= \E(y_{ijr^\star})$ of the predictive
distribution for computing
the DSS and the mean squared error score. Using the law of iterated
expectations [\citet{Billingsley-1986}, Theorem 34.4],
$\mu_{ijr^\star}$ can be derived. With $y_{ijr^\star}|\lambda
_{ijr^\star
} \sim\po(n_{ijr^\star}\cdot\lambda_{ijr^\star})$, it follows
\[
\mu_{ijr^\star} = \E(\E(y_{ijr^\star}|\lambda_{ijr^\star})) = \E
(n_{ijr^\star}\cdot\lambda_{ijr^\star}) = n_{ijr^\star} \cdot\E
(\lambda_{ijr^\star}).
\]
Analogously, the variance $\sigma^2_{ijr^\star}= \Var(y_{ijr^\star})$
follows from the law of total variance as
\begin{eqnarray*}
\sigma^2_{ijr^\star} &=& \E(\Var(y_{ijr^\star}|\lambda_{ijr^\star
})) +
\Var(\E(y_{ijr^\star}|\lambda_{ijr^\star}))\\
&=& \E(n_{ijr^\star}\cdot\lambda_{ijr^\star}) + \Var(n_{ijr^\star
}\cdot
\lambda_{ijr^\star}) \\
&=& n_{ijr^\star} \cdot\E(\lambda_{ijr^\star
}) +
n_{ijr^\star}^2 \Var(\lambda_{ijr^\star})
\end{eqnarray*}
for INLA. Under a quasi-Poisson approach with $\Var(y_{ijr}|\lambda
_{ijr}) = \phi\cdot n_{ijr} \cdot\lambda_{ijr}$
we need to explicitly incorporate the overdispersion parameter $\phi$, so
$\sigma^2_{ijr} = \phi\cdot n_{ijr} \E(\lambda_{ijr}) + n_{ijr}^2
\Var
(\lambda_{ijr})$.
Here, we used the total lack of fit as $\phi$; compare
\citet{booth-etal-2002}.

To obtain posterior predictive quantiles, the missing Poisson variation
was added
to the predicted mortality rates. When using INLA, this is done by
numerical integration over
the predicted posterior marginal of $\lambda_{ijr^\star}$. Since we do
not obtain the posterior marginal
of $\lambda_{ijr^\star}$ using the ilc-package, we used Monte Carlo
sampling instead.
To be more precise, we generated $N=100\mbox{,}000$ samples for the
linear predictor $\eta_{ijr^\star}$
from a normal distribution with mean $\log(\hat{\lambda}_{ijr^\star})$
and variance derived from the
symmetric prediction intervals on log-scale. Then, we generated for
each sample $\eta_{ijr^\star}^{(s)}$,
$s=1, \ldots, N$, one sample $y_{ijr^\star}^{(s)}$ from a negative
binomial distribution with density
\[
f(y) = \frac{\Gamma(y+d)}{\Gamma(d)\Gamma(y+1)} \biggl(\frac
{d}{m+d}\biggr)^d \biggl(\frac{m}{m+d}\biggr)^y,
\]
where $\E(y) = m$ and $\Var(y) = m( 1+m/d)$. To match the mean and
variance of the quasi-Poisson distribution,
we set $m^{(s)}= n_{ijr^\star}\exp(\eta^{(s)}_{ijr^\star})$ and
$d^{(s)}=m^{(s)}/(\phi-1)$ for each
sample $\eta_{ijr^\star}^{(s)}$.
Subsequently, quantiles at different prediction levels could be
extracted from the samples.

%
%
\begin{table}
\tabcolsep=0pt
\caption{Mean Dawid--Sebastiani score ($\overline{\mathit{DSS}}$), mean
squared error score (MSE),
empirical coverage probabilities for all models predicting female
mortality of one country either
for the first or second half of the 20th century} \label{tabpredAss}
\begin{tabular*}{\tablewidth}{@{\extracolsep{\fill}}rccccccc@{}}
\hline
&&\multicolumn{3}{c}{\textbf{1900--1949}} & \multicolumn{3}{c@{}}{\textbf{1950--1999}}\\[-4pt]
&&\multicolumn{3}{c}{\hrulefill} & \multicolumn{3}{c@{}}{\hrulefill}\\
&\textbf{Measure}& \textbf{Lee--Carter} & \textbf{APC} & \textbf{cMAPC}
& \textbf{Lee--Carter} & \textbf{APC} & \textbf{cMAPC} \\
\hline
&&\multicolumn{6}{c@{}}{{NORWAY}}\\[4pt]
&$\overline{\mathrm{DSS}}$ & 232.4 & 38.9 & 17.9 & 13.5 & 16.8 & 15.0
\\
&MSE & 3.49e$+$06 & 7.21e$+$06 & 1.73e$+$06 & 3.10e$+$06 & 1.92e$+$07 & 1.50e$+$07
\\[3pt]
Level
&95\% & 19 & 66 & 70 & 91
& 96 & 78 \\
&80\% & 13 & 52 & 35 & 79 & 73 & 54 \\
&50\% & \hphantom{0}9 & 39 & 11 & 52 & 35 & 37 \\
[4pt]
&&\multicolumn{6}{c}{{DENMARK}}\\
[4pt]
&$\overline{\mathrm{DSS}}$ & 50.5 & 41.2 & 14.7 & 22.2 & 17.1 & 13.7
\\
&MSE & 3.37e$+$06 & 2.30e$+$07 & 3.56e$+$06 & 2.03e$+$07 & 6.43e$+$07 & 1.80e$+$07
\\[3pt]
Level
&95\% & 24 & 58 & 96 & 66
& 95 & 99 \\
&80\% & 17 & 50 & 93 & 51 & 92 & 83 \\
&50\% & \hphantom{0}6 & 15 & 79 & 30 & 64 & 65 \\
[4pt]
&&\multicolumn{6}{c}{{SWEDEN}}\\
[4pt]
&$\overline{\mathrm{DSS}}$ & 249.6 & 43.4 & 15.4 & 24.7 & 18.5 & 13.6
\\
&MSE & 4.45e$+$07 & 9.48e$+$07 & 9.04e$+$06 & 9.43e$+$07 & 2.04e$+$08 & 3.05e$+$06
\\[3pt]
Level
&95\% & 11 & 64 & 99 & 70
& 97 & 95 \\
&80\% & \hphantom{0}8 & 28 & 94 & 62 & 95 & 72 \\
&50\% & \hphantom{0}4 & \hphantom{0}8 & 55 & 41 & 75 & 54 \\
\hline
\end{tabular*}
\end{table}

Table \ref{tabpredAss} shows for all models the mean squared error
(MSE) score,
the empirical coverage probabilities and the mean DSS
averaged over all 170 projections.
For five of the six scenarios and especially when predicting the first
10 periods,
the correlated multivariate APC model is clearly the best model.
Although the prediction intervals
are sometimes too large, as indicated by larger empirical coverage
probabilities than the nominal level,
the empirical coverage is mostly closer to the nominal level than for
the other two approaches.
Regarding mean DSS and empirical coverage, the univariate APC model
also performs mostly better than the
extended Lee--Carter approach. In particular, predicting the first half
of the 20th century, the extended
Lee--Carter approach showed severe deficits. It was classified as the
best model regarding all predictive
assessment criteria only when predicting the second half of the 20th century
for Norway. Inspecting the posterior correlation estimates of the cMAPC model
(Table \ref{tabcorSum}), we observe that for this scenario the
posterior correlations
between country-specific period effects and also between country-specific
cohort effects are lower than in the other scenarios. In contrast, the
correlation
between country-specific overdispersion parameters is quite high.

%
%
\begin{table}
\caption{Median and $95\%$ credible interval for all correlation
parameters in the correlated multivariate APC~model} \label{tabcorSum}
\begin{tabular*}{\tablewidth}{@{\extracolsep{\fill}}lccccc@{}}
\hline
\textbf{Predicted} & \textbf{Predicted} &&&&\\
\textbf{period} & \textbf{country} & \textbf{Age} & \textbf{Period}
& \textbf{Cohort} & \textbf{Overdispersion}\\
\hline
1900--1949 &Norway & $_{{0.978}} {0.991}_{ {0.996}}$ & $_{{0.81}} {0.96}_{
{0.99}}$ & $_{{0.57}} {0.82}_{ {0.94}}$ & $_{{0.91}} {0.93}_{ {0.94}}$
\\[2pt]
&Denmark & $_{{0.994}} {0.998}_{ {0.999}}$ & $_{{0.35}}
{0.87}_{ {0.99}}$ & $_{{0.44}} {0.76}_{ {0.92}}$ & $_{{0.83}} {0.86}_{
{0.89}}$ \\[2pt]
&Sweden & $_{{0.984}} {0.994}_{ {0.997}}$ & $_{{0.70}} {0.95}_{
{0.99}}$ & $_{{0.58}} {0.84}_{ {0.95}}$ & $_{{0.75}} {0.79}_{ {0.84}}$
\\[6pt]
1950--1999 &Norway & $_{{0.969}} {0.986}_{ {0.994}}$ & $_{{0.01}} {0.51}_{
{0.85}}$ & $_{{0.28}} {0.62}_{ {0.84}}$ & $_{{0.90}} {0.92}_{ {0.94}}$
\\[2pt]
&Denmark & $_{{0.964}} {0.984}_{ {0.994}}$ & $_{{0.18}}
{0.75}_{ {0.96}}$ & $_{{0.89}} {0.97}_{ {0.99}}$ & $_{{0.82}} {0.85}_{
{0.88}}$ \\[2pt]
&Sweden & $_{{0.976}} {0.990}_{ {0.996}}$ & $_{{0.62}} {0.96}_{
{1.00}}$ & $_{{0.60}} {0.83}_{ {0.94}}$ & $_{{0.77}} {0.81}_{ {0.84}}$
\\
\hline
\end{tabular*}
\end{table}

\begin{figure}

\includegraphics{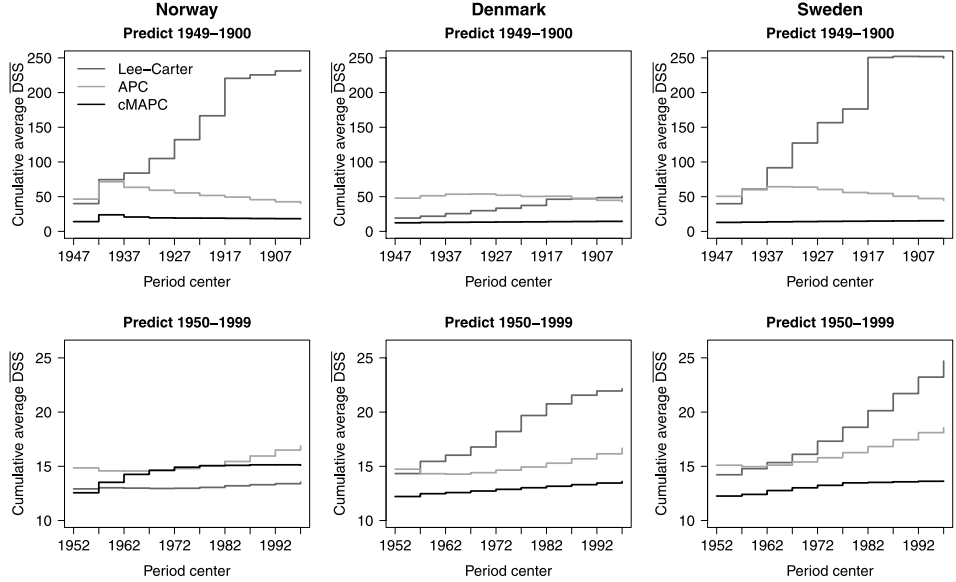}

\caption{Cumulative average of mean Dawid--Sebastiani scores across age
groups.}\label{figcumDSS}
\end{figure}

To compare the performance change from short-term to long-term forecasts,
Figure \ref{figcumDSS} shows the cumulative average $\overline
{\mathrm{DSS}}_j$,
where $\overline{\mathrm{DSS}}_j$ denotes the mean DSS across age group
at period $j$.
%
%
Except for predicting death rates in Norway from 1950--1999, the curve
for the cMAPC
model is always below those of the two univariate approaches.
Predicting the periods in the first
half of the 20th century, the cumulative average $\overline{\mathrm
{DSS}}_j$ of the extended Lee--Carter model
is constantly increasing, indicating a lower projection quality with
increasing time. The largest jump occurs
for the period 1915--1919 with the Spanish flu.
In contrast, the score of the univariate APC model decreases when
predicting more periods, while
the score for the correlated multivariate APC model stays fairly constant.
Predicting the periods in the second half of the 20th century, the
cumulative average $\overline{\mathrm{DSS}}_j$
slowly increases for all models. However, except for Norway, the
cumulative score of the
extended Lee--Carter model shows larger jumps from one period to the
next.

\subsubsection{Projections}
\begin{sidewaysfigure}
%
%

\includegraphics{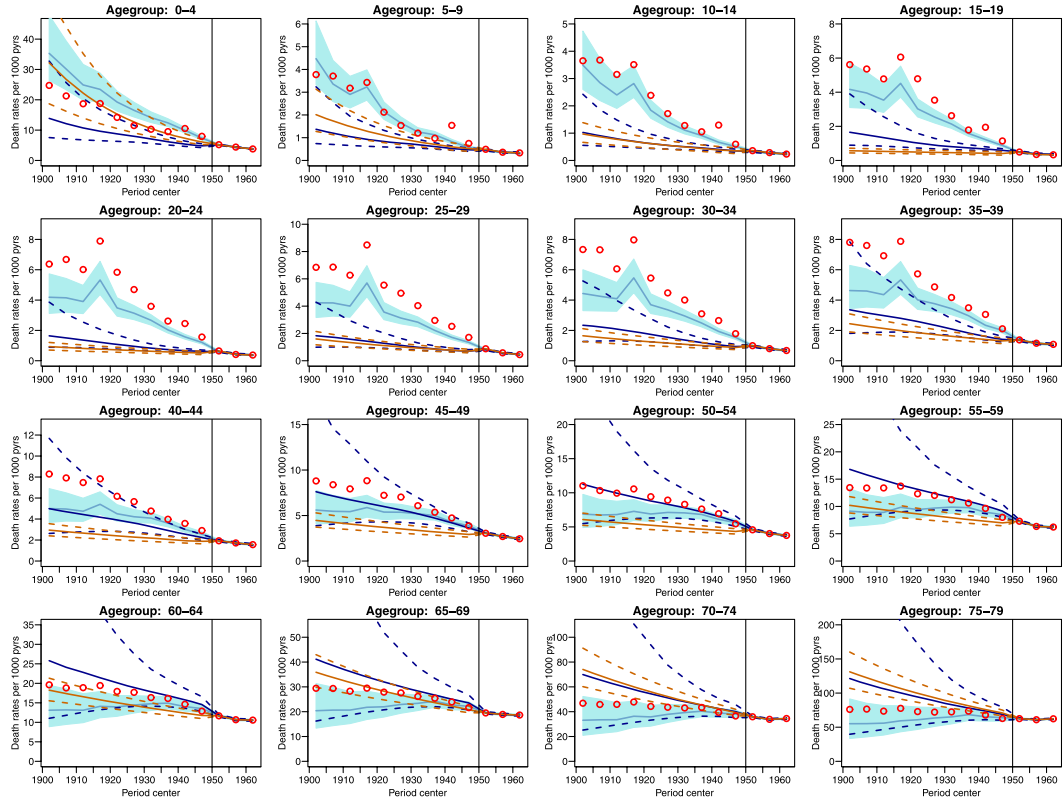}\centering

\caption{Median predicted mortality rates per 1\mbox{,}000 person-years (pyrs)
of Norwegian women
within 80\% CI regions for the years 1900--1949. Shown are the results
from an
extended Lee--Carter model (orange lines), a~univariate APC model (dark
blue lines)
and a~multivariate APC model with age, period, cohort and
overdispersion parameters
correlated across regions (light blue shaded). In addition, the true
mortality rates
for Norwegian women (red $\circ$) are shown.}
\label{fign1}
\end{sidewaysfigure}

\begin{sidewaysfigure}\centering
%

\includegraphics{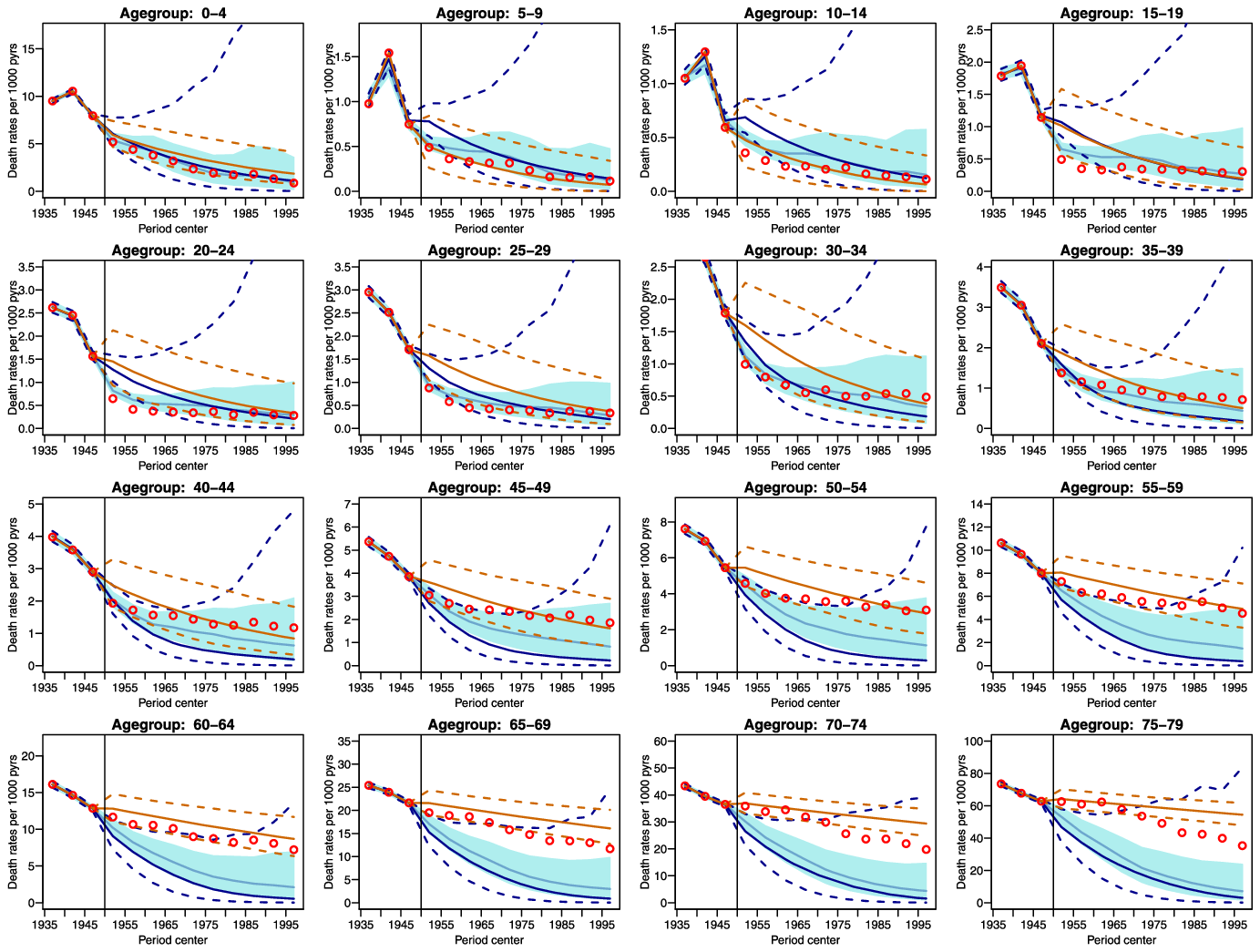}

\caption{Median predicted mortality rates per 1\mbox{,}000 person-years (pyrs)
of Norwegian women
within 80\% CI regions for the years 1950--1999. Shown are the results
from an
extended Lee--Carter model (orange lines), a univariate APC model (dark
blue lines)
and a~multivariate APC model with age, period, cohort and
overdispersion parameters
correlated across regions (light blue shaded). In addition, the true
mortality rates
for Norwegian women (red $\circ$) are shown.} \label{fign2}
\end{sidewaysfigure}

The median projected death rates per $1\mbox{,}000$ person-years
together with 80\% pointwise prediction intervals
for Norwegian women obtained from all three models are shown in
Figure \ref{fign1} for the first half
and Figure \ref{fign2} for the second half of the 20th century.
Furthermore, the true death rates of Norwegian
women are added to the figures.
For all models and especially for the univariate APC model, the
prediction intervals are getting wider
as prediction time goes on. While the projections of the two univariate
approaches are almost straight lines,
different temporal patterns across age groups can be seen for the
correlated model. The Spanish flu
was especially well captured by the correlated approach. Since the
Spanish flu did not affect all
age groups, this event (also present for Denmark and Sweden) was
captured and transferred to the projections
due to the correlated overdispersion parameters and not because of the
correlated period effects,
as one might intuitively guess. Predicting the second half of the 20th
century, the projections of the extended
Lee--Carter model agree very well with the true observations. In
particular, the rates for ages over 60
are well projected, where the cMAPC model tends to underestimate the
overall death rate.\looseness=1

Projections for Danish and Swedish women are shown in Figures \ref
{figdk1}, \ref{figdk2},
\ref{figsv1} and~\ref{figsv2} in Appendix \ref{secappendixB}.
Here, the projections of the
correlated approach coincide for all scenarios and all age groups very
well with the observed rates.
In contrast, the extended Lee--Carter approach tends to underestimate
rates for younger age groups when predicting the
first half, and to overestimate rates for older age groups when
predicting the second half of the 20th century.\looseness=1

\section{Discussion}\label{secdiscussion}
In this paper we proposed the use of correlated smoothing priors and
correlated overdispersion parameters
for multivariate Bayesian APC approaches analyzing mortality or
morbidity rates stratified by age, period, cohort and one further variate.
The specification of correlated smoothing priors involves a Kronecker
product precision structure for the outcome-specific
time effects, that is, age, period and/or cohort effects. We
implemented correlated multivariate APC models based on
a uniform correlation structure in MCMC and INLA. In the first
application we analyzed COPD mortality
among males in England and Wales using MCMC and INLA, and compared the
results of an ordinary multivariate
APC model with those obtained from different correlated model
formulations. A comparison of MCMC
and INLA showed virtually identical results. As indicated by the log
marginal likelihood, the formulation
with both correlated overdispersion and correlated stratum-specific
period and cohort effects was classified
as the best. As shown in the relative risk estimates, the correlated
model structure improved the precision of the relative risk
estimates especially for younger birth cohorts.

In a second application on overall mortality of Danish, Swedish and
Norwegian women in the 20th century,
we performed a cross-prediction study. We illustrated the good
predictive quality of the correlated
approach when imputing missing data units for one country if these
units are available for the other countries.
As focus was set on projections, which are an estimable function in
(multivariate) APC models, we were able to consider the most flexible
model with country-specific age, period, cohort and overdispersion
parameters that were all correlated across
countries. In total, we considered six scenarios treating in turn for a
particular country either the first
or second half of the data as missing and subsequently predicting the
omitted data units.
We compared the projections to those obtained from a~univariate APC model
and a Lee--Carter approach embedded into a quasi-Poisson
model using the proper Dawid--Sebastiani scoring rule. Since only the
correlated formulation can take advantage
from the complete tables of the remaining two countries, it was
classified as
the best model in five of the six scenarios. Furthermore, we observed
that the predictive quality stayed almost
constant when increasing the number of periods to predict, which was
not the case for the two
univariate approaches. Thus, the correlated approach outperformed both
univariate methods
in short and long-term projections.\looseness=1

In real life, long-term projections of mortality or disease rates into the
future are difficult to make using our proposed approach, since there
will be no data from comparable
time-series available. For short-term predictions, data for some strata
may be already available, while for
others they are still missing. Here, the correlation approach will be
useful to forecast the missing units.

For the simultaneous projection of several strata, \citet{li-lee-2005}
extended the original log-linear Lee--Carter model.
In the simplest extension, they assigned each stratum its own age
pattern, while assuming shared age-specific patterns
of mortality change and a shared time trend. Future values are then
predicted for
the shared time trend based on an ARIMA process. Incorporating both
shared and stratum-specific parameters, this model
seems to be similar in spirit to an uncorrelated multivariate APC model
[\citet{riebler-held-2010}].
In a more complex model, \citet{li-lee-2005} included an additional
stratum-specific bilinear term
to allow for differences between the rate of change in mortality in a
particular stratum and the rate of change implied
by the common bilinear term. However, in contrast to the correlated
approach presented here, those extensions cannot take advantage of
data units missing in one stratum but present for the remaining. By
contrast, our approach could be equally used
to impute data for all strata simultaneously, benefitting from the
periods where complete data existed.
Furthermore, we can use stratum-specific effects for all parameters.
Information from the remaining strata is
borrowed by incorporating correlation. In a Bayesian setting the
inclusion of correlation between parameters is
straightforward via the prior distributions, whereas in a frequentist
setting this seems to be more complex.

Another interesting field of application is similar in spirit to the
inference on collapsed margins,
proposed by \citet{byers-besag-2000}. In the context of collapsed
margins, complete data are available
on several risk factors, but a subsequent analysis indicates that
information on an additional
variable is relevant. For this variable the numbers of persons at risk
are available but
not the numbers of cases. \citet{byers-besag-2000} propose a Bayesian
approach to estimate the
effect of the variable. In multivariate APC models it might be that
multiple data sets are
only available for a specific period in time, while, before and/or after
this date, data only exist for the conjunction of outcomes. A typical
example could be Germany,
which was formerly united, then separated and now united again. Using
age-specific data on the population
sizes from 1990 up to now for East and West Germany separately, it may
be possible to project mortality rates for both
individual parts, by exploiting the correlation present when they were
divided. Thus, the observations
for the conjunction of both parts could be separated.
However, further investigations are required to explore the applicability.

The proposed methodology can only be applied to data stratified by one further
variate. For analyzing mortality rates stratified by more than one further
variate, a conditional approach using a multinomial logistic regression
model has
been proposed [\citet{held-riebler-2010}]. However, the
incorporation of
correlation has not yet
been considered.

A disadvantage of the proposed methodology might be that it is essentially
additive in age, period and cohort, so that
interactions between the time dimensions cannot be explicitly modeled.
\citet{currie-etal-2004}
proposed two-dimensional smoothing to address this problem in the
analysis of an individual
registry data set. \citet{biatat-currie-2010} started to extend this
work and proposed a model
to compare various mortality tables by assuming a common
two-dimensional P-spline surface and
additional one-dimensional smoothing functions for age and period.

In general, the use of a Kronecker product structure is a promising
area for further research, as different correlation structures can easily
be combined with different precision matrices. Based on the uniform
correlation structure INLA can, by now,
correlate a wide range of other GMRF models as components of more
general additive regression models.
Examples are as follows: nonparametric seasonal models, continuous-time
random walks or models with a user
specified precision matrix. However, the uniform correlation structure
is rather restrictive and may only be plausible for a few outcomes.
Future work encompasses the integration
of other correlation structures, for example, depending on the distance
between units, so that the approach
can be extended to the space--time context, for example. Furthermore, we
are investigating
the use of correlated two-dimensional smoothing priors in INLA to
incorporate interactions
between time-dimensions into the multivariate APC model.\vspace*{6pt}

\begin{appendix}
\section{Uniform correlation structure} \label{secappendix}\vspace*{3pt}

Let ${\mathbf{C}}$ be an $R\times R$ correlation matrix with
uniform correlation structure, so that ${\mathbf{C}} = (1-\rho
){\mathbf
{I}} + \rho\mathbf{J}$:
\[
{\mathbf{C}} =
\pmatrix{
1 & \rho& \cdots& \rho\vspace*{1pt}\cr
\rho& \ddots& \ddots&\vdots\vspace*{1pt}\cr
\vdots& \ddots& \ddots& \rho\vspace*{1pt}\cr
\rho& \cdots& \rho& 1},\vspace*{2pt}
\]
where $\rho$ is the correlation parameter, $\mathbf{I}$ denotes the
$R\times R$ identity matrix
and $\mathbf{J}$ an $R \times R$ matrix of ones. Then the inverse
$\mathbf{C}^{-1}$ is given by
\[
\mathbf{C}^{-1} =
\pmatrix{
a & b & \cdots& b \vspace*{1pt}\cr
b & \ddots& \ddots&\vdots\vspace*{1pt}\cr
\vdots& \ddots& \ddots& b\vspace*{1pt}\cr
b& \cdots& b & a}\vspace*{2pt}
\]
with
\begin{eqnarray*}
a &=& -\frac{(R-2)\cdot\rho+1}{(\rho-1)\{(R-1)\cdot\rho+1\}},\\
b &=& \frac{\rho}{(\rho-1)\{(R-1)\cdot\rho+1\}}.
\end{eqnarray*}
\begin{pf}
If $\mathbf{C}^{-1} \mathbf{C} = \mathbf{I}$, then $\mathbf
{C}^{-1}$ is
the inverse of $\mathbf{C}$.
For the diagonal elements of $\mathbf{C}^{-1} \mathbf{C}$ it follows that
\begin{eqnarray*}
(\mathbf{C}^{-1} \mathbf{C})_{(i,i)} &=& a + (R-1)\cdot b\cdot\rho\\
&=& \frac{-(R-2)\cdot\rho-1 + (R-1)\cdot\rho^2}{(\rho-1)\{
(R-1)\cdot\rho+1\}} \\
&=& \frac{-R\rho+ 2\rho- 1 + R\rho^2 - \rho^2}{R\rho^2-\rho^2 -
R\rho+ \rho+\rho- 1} = 1
\end{eqnarray*}
for all $i=1, \ldots, R$. For the nondiagonal elements, that is, $i\ne
j$, we get
\begin{eqnarray*}
(\mathbf{C}^{-1} \mathbf{C})_{(i,j)} & = & a \cdot\rho+ b +
(R-2)\cdot b \cdot\rho\\
&=& \frac{\{-(R-2)\cdot\rho-1\}\rho+ \rho+ (R-2)\cdot\rho
^2}{(\rho
-1)\{(R-1)\cdot\rho+1\}} \\
&=& \frac{-R\rho^2 + 2\rho^2 - \rho+ \rho+R\rho^2 - 2\rho
^2}{(\rho
-1)\{(R-1)\cdot\rho+1\}} = 0.
\end{eqnarray*}
\upqed
\end{pf}

The determinant $|\mathbf{C}^{-1}|$ is given by
\[
|\mathbf{C}^{-1}| = |\mathbf{C}|^{-1} =
\bigl[\bigl(1+(R-1)\rho\bigr)(1-\rho)^{R-1}\bigr]^{-1}.
\]
\begin{pf}
We show that $|\mathbf{C}|=(1+(R-1)\rho)(1-\rho)^{R-1}$,
as the inverse case follows immediately. Remember that
$|\mathbf{C}| = |\mathbf{I} - \rho\mathbf{I} + \rho\mathbf{J}|$.
The identity matrix has $R$ times the eigenvalue $1$. The matrix
$\mathbf{J}$ has once the eigenvalue~$R$ and $R-1$ times the
eigenvalue $0$. Since both matrices ($\mathbf{I}$ and $\mathbf{J}$)
share the same eigenvectors,
the eigenvalues for $\mathbf{C}$ are $(1-\rho+ \rho\cdot R)$
and $(1-\rho)$ with multiplicity $R-1$, so that
the determinant of $\mathbf{C}$, the product of the eigenvalues, is
\[
|\mathbf{C}| = (1 - \rho+ \rho\cdot R)(1-\rho)^{R-1} = \bigl(1 +
(R-1)\rho
\bigr)(1-\rho)^{R-1}.
\]
\upqed
\end{pf}

%
\section{Projection for Denmark and Sweden}\label{secappendixB}
\begin{sidewaysfigure}\centering
%

\includegraphics{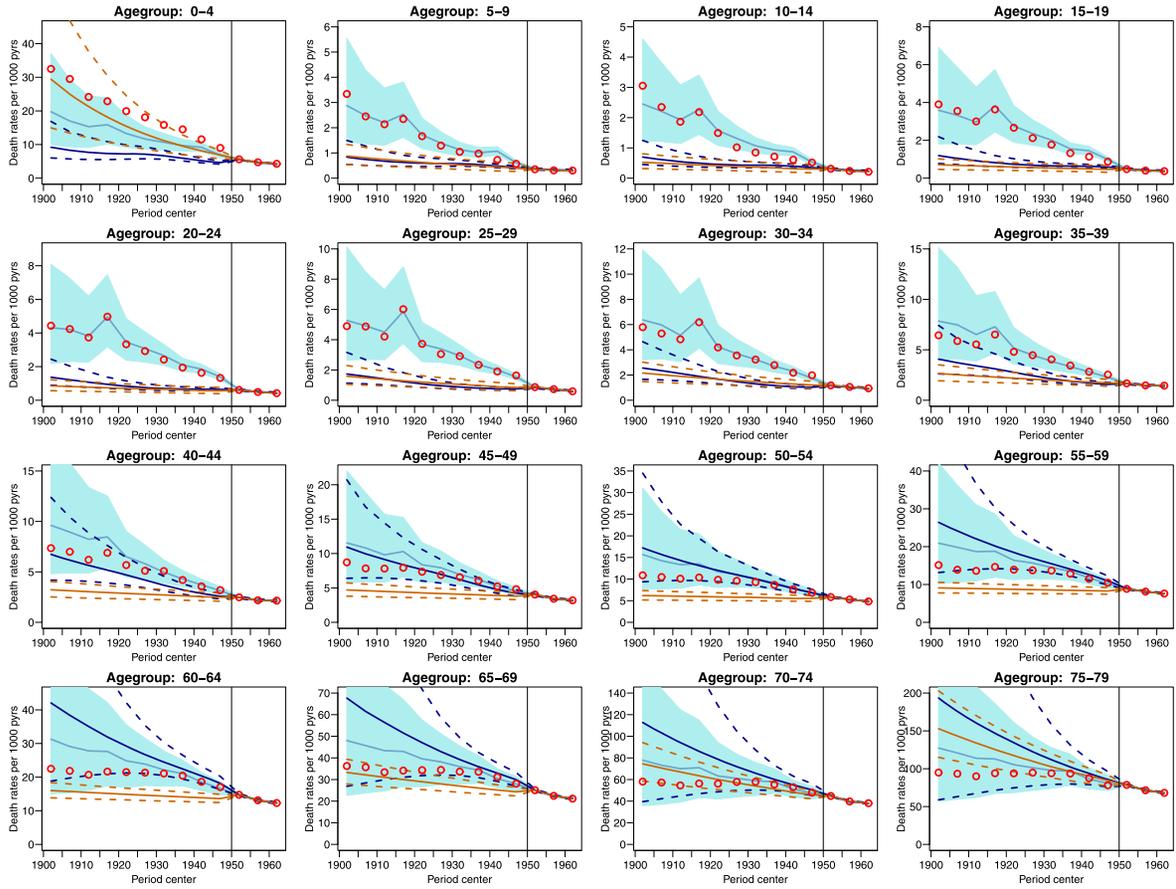}

\caption{Same quantities as in Figure \protect\ref{fign1} but for
Danish women (red $\circ$).}\label{figdk1}
\end{sidewaysfigure}

\begin{sidewaysfigure}\centering
%

\includegraphics{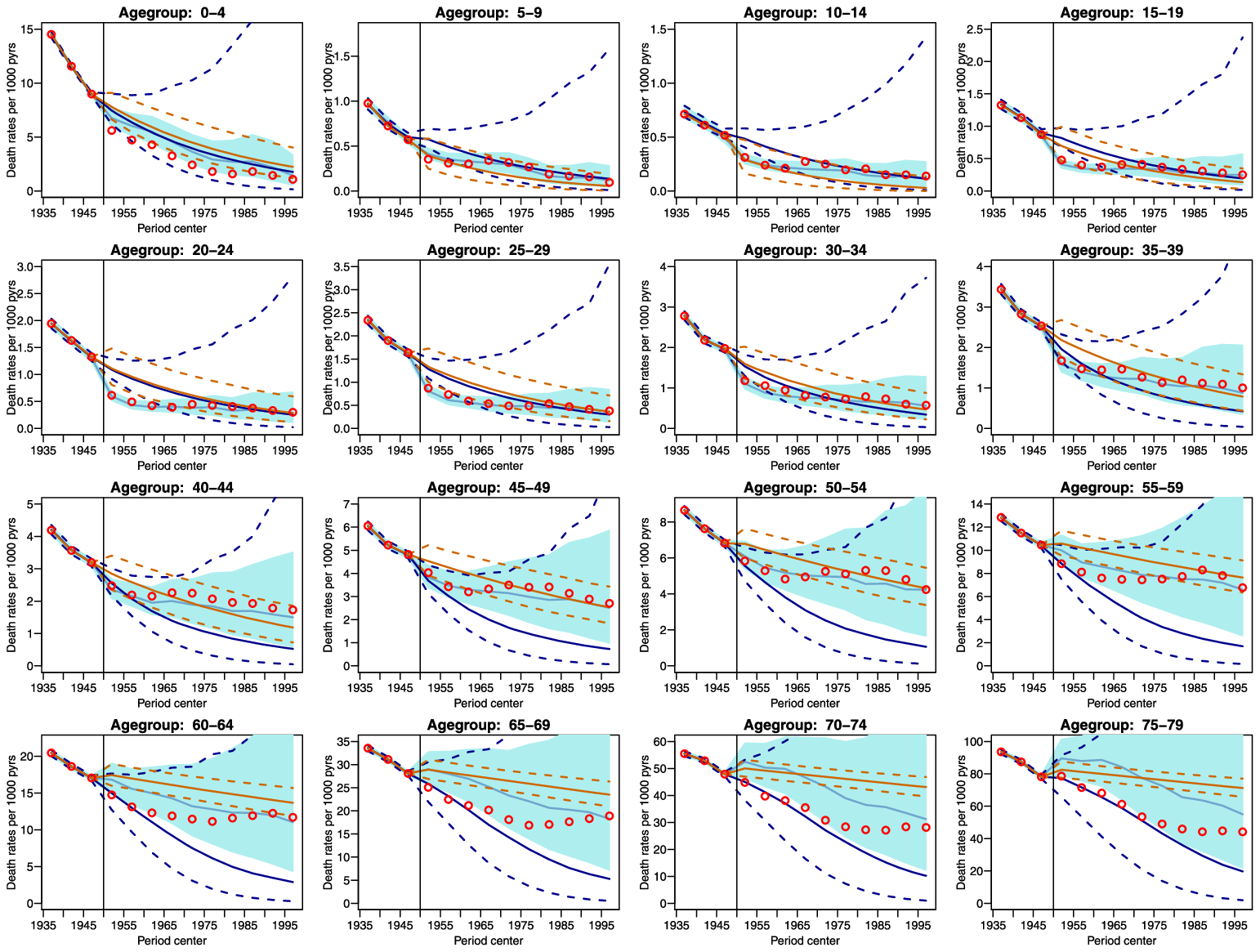}

\caption{Same quantities as in Figure \protect\ref{fign2} but for
Danish women (red $\circ$).}\label{figdk2}
\end{sidewaysfigure}

\begin{sidewaysfigure}\centering
%

\includegraphics{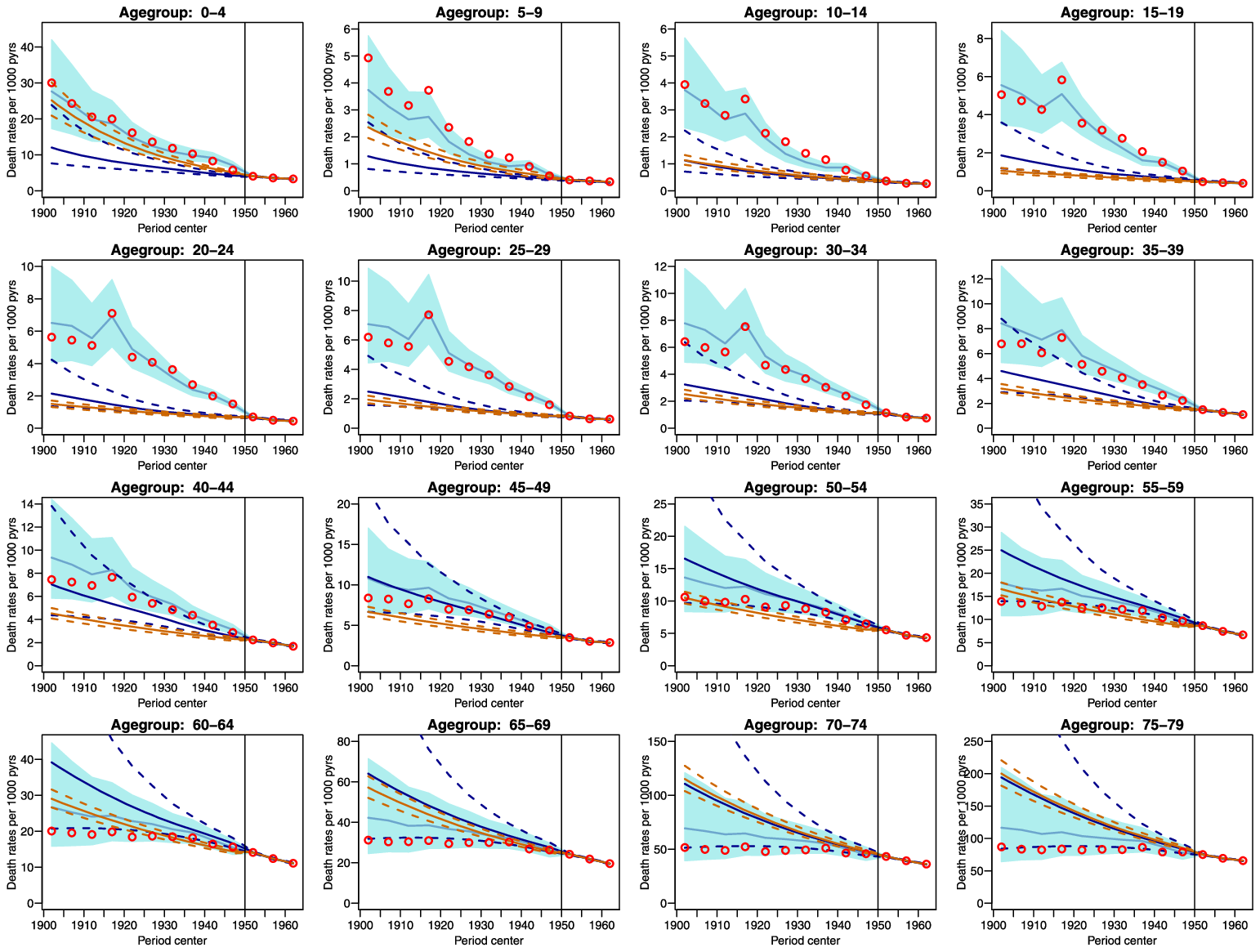}

\caption{Same quantities as in Figure \protect\ref{fign1} but for
Swedish women (red $\circ$).}\label{figsv1}
\end{sidewaysfigure}

\begin{sidewaysfigure}\centering
%

\includegraphics{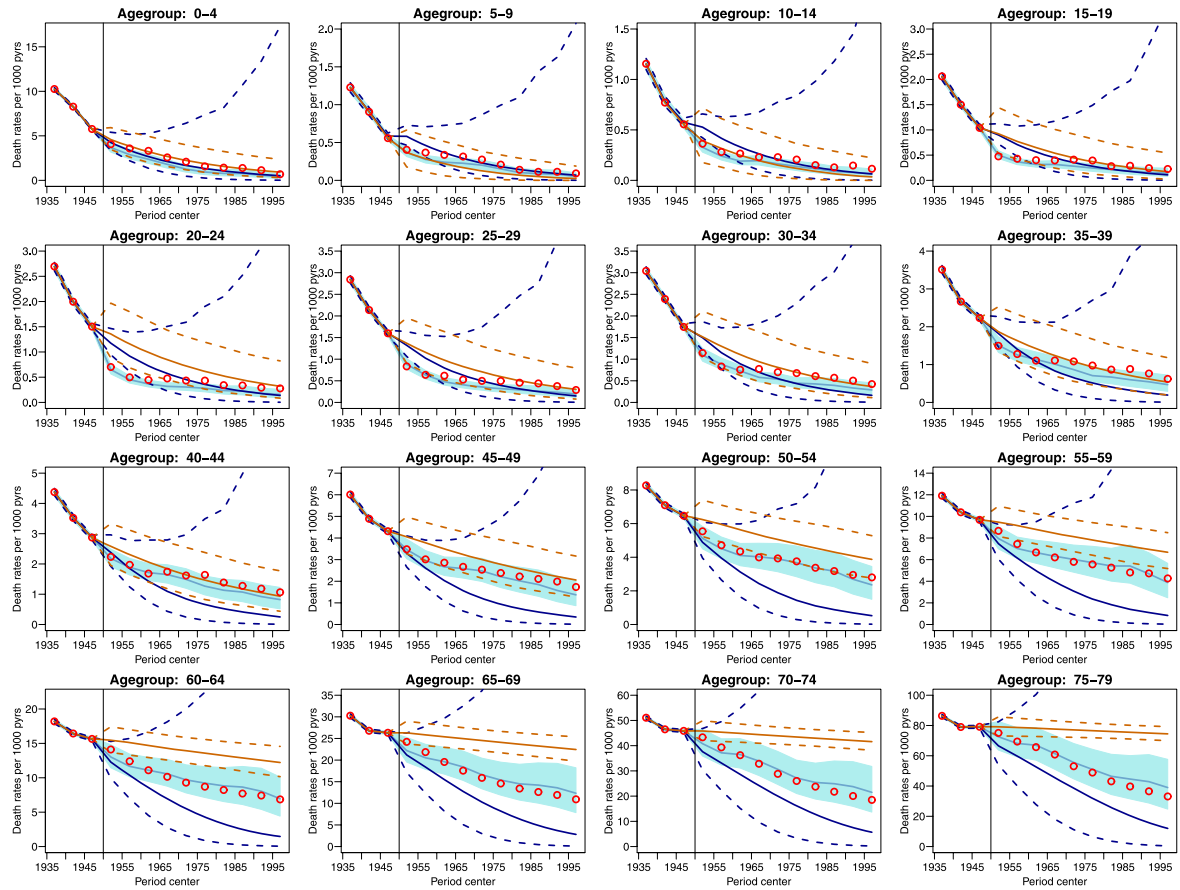}

\caption{Same quantities as in Figure \protect\ref{fign2} but for
Swedish women (red $\circ$).}\label{figsv2}
\end{sidewaysfigure}

The median projected death rates per $1\mbox{,}000$ person-years
together with 80\% pointwise prediction intervals
for Danish and Swedish women obtained from all three models are shown
in Figures \ref{figdk1} and \ref{figsv1}
for the first half and Figures \ref{figdk2} and \ref{figsv2} for
the second half of the 20th century.

\end{appendix}

\section*{Acknowledgments}

We would like to thank the Editor and referees for their helpful
comments and suggestions that improved the paper.

\begin{supplement}[id=suppA]\label{suppA}
\stitle{Code repository for the cross-prediction study of overall
mortality of Scandinavian women}
\slink[doi]{10.1214/11-AOAS498SUPP} 
\slink[url]{http://lib.stat.cmu.edu/aoas/498/supplement.zip}
\sdatatype{.zip}
\sdescription{This repository archives the data,
R-code and results for the cross-prediction study of overall mortality
of Scandinavian women presented in Section \ref{secscan}. In
particular, it contains code to make Table \ref{tabpredAss} and
Figures \ref{figcumDSS}--\ref{figsv2}.}
\end{supplement}

%

\printaddresses


\begin{thebibliography}{74}

\bibitem[\protect\citeauthoryear{Andreasen, Viboud and
  Simonsen}{2008}]{andreasen-etal-2008}
\begin{barticle}[pbm]
\bauthor{\bsnm{Andreasen},~\bfnm{Viggo}\binits{V.}},
  \bauthor{\bsnm{Viboud},~\bfnm{C{\'{e}}cile}\binits{C.}} \AND
  \bauthor{\bsnm{Simonsen},~\bfnm{Lone}\binits{L.}}
(\byear{2008}).
\btitle{Epidemiologic characterization of the 1918 influenza pandemic summer
  wave in Copenhagen: Implications for pandemic control strategies}.
\bjournal{J. Infect. Dis.}
\bvolume{197}
\bpages{270--278}.
\bid{doi={10.1086/524065}, issn={0022-1899}, mid={NIHMS96642}, pmcid={2674012},
  pmid={18194088}}
\bptok{imsref}%
\end{barticle}
\endbibitem

\bibitem[\protect\citeauthoryear{Armitage}{1966}]{armitage-1966}
\begin{barticle}[mr]
\bauthor{\bsnm{Armitage},~\bfnm{P.}\binits{P.}}
(\byear{1966}).
\btitle{The chi-square test for heterogeneity of proportions, after adjustment
  for stratification}.
\bjournal{J. Roy. Statist. Soc. Ser. B}
\bvolume{28}
\bpages{150--163}.
\bid{issn={0035-9246}, mr={0196851}}
\bptok{imsref}%
\end{barticle}
\endbibitem

\bibitem[\protect\citeauthoryear{Baker and Bray}{2005}]{baker-bray-2005}
\begin{barticle}[pbm]
\bauthor{\bsnm{Baker},~\bfnm{A.}\binits{A.}} \AND
  \bauthor{\bsnm{Bray},~\bfnm{I.}\binits{I.}}
(\byear{2005}).
\btitle{Bayesian projections: What are the effects of excluding data from
  younger age groups?}
\bjournal{Am. J. Epidemiol.}
\bvolume{162}
\bpages{798--805}.
\bid{doi={10.1093/aje/kwi273}, issn={0002-9262}, pii={kwi273}, pmid={16135506}}
\bptok{imsref}%
\end{barticle}
\endbibitem

\bibitem[\protect\citeauthoryear{Berzuini and
  Clayton}{1994}]{berzuini-clayton-1994}
\begin{barticle}[pbm]
\bauthor{\bsnm{Berzuini},~\bfnm{C.}\binits{C.}} \AND
  \bauthor{\bsnm{Clayton},~\bfnm{D.}\binits{D.}}
(\byear{1994}).
\btitle{Bayesian analysis of survival on multiple time scales}.
\bjournal{Stat. Med.}
\bvolume{13}
\bpages{823--838}.
\bid{issn={0277-6715}, pmid={8047738}}
\bptok{imsref}%
\end{barticle}
\endbibitem

\bibitem[\protect\citeauthoryear{Besag et~al.}{1995}]{besag-etal-1995}
\begin{barticle}[mr]
\bauthor{\bsnm{Besag},~\bfnm{Julian}\binits{J.}},
  \bauthor{\bsnm{Green},~\bfnm{Peter}\binits{P.}},
  \bauthor{\bsnm{Higdon},~\bfnm{David}\binits{D.}} \AND
  \bauthor{\bsnm{Mengersen},~\bfnm{Kerrie}\binits{K.}}
(\byear{1995}).
\btitle{Bayesian computation and stochastic systems}.
\bjournal{Statist. Sci.}
\bvolume{10}
\bpages{3--66}.
\bid{issn={0883-4237}, mr={1349818}}
\bptnote{check related}%
\bptok{imsref}%
\end{barticle}
\endbibitem

\bibitem[\protect\citeauthoryear{Biatat and Currie}{2010}]{biatat-currie-2010}
\begin{binproceedings}[author]
\bauthor{\bsnm{Biatat},~\bfnm{V~D}\binits{V.~D.}} \AND
  \bauthor{\bsnm{Currie},~\bfnm{I~D}\binits{I.~D.}}
(\byear{2010}).
\btitle{Joint models for classification and comparison of mortality in
  different countries}.
In \bbooktitle{25th International Workshop on Statistical Modelling}
(\beditor{\bfnm{A~W}\binits{A.~W.}~\bsnm{Bowman}}, ed.)
\bpages{89--94}.
\bpublisher{Univ. Glasgow}, \baddress{UK}.
\bptok{imsref}%
\end{binproceedings}
\endbibitem

\bibitem[\protect\citeauthoryear{Billingsley}{1986}]{Billingsley-1986}
\begin{bbook}[mr]
\bauthor{\bsnm{Billingsley},~\bfnm{Patrick}\binits{P.}}
(\byear{1986}).
\btitle{Probability and Measure},
\bedition{2nd} ed.
\bpublisher{Wiley}, \baddress{New York}.
\bid{mr={0830424}}
\bptok{imsref}%
\end{bbook}
\endbibitem

\bibitem[\protect\citeauthoryear{Booth}{2006}]{booth-2006}
\begin{barticle}[author]
\bauthor{\bsnm{Booth},~\bfnm{H}\binits{H.}}
(\byear{2006}).
\btitle{Demographic forecasting: 1980 to 2005 in review}.
\bjournal{International Journal of Forecasting}
\bvolume{22}
\bpages{547--581}.
\bptok{imsref}%
\end{barticle}
\endbibitem

\bibitem[\protect\citeauthoryear{Booth, Maindonald and
  Smith}{2002}]{booth-etal-2002}
\begin{barticle}[pbm]
\bauthor{\bsnm{Booth},~\bfnm{Heather}\binits{H.}},
  \bauthor{\bsnm{Maindonald},~\bfnm{John}\binits{J.}} \AND
  \bauthor{\bsnm{Smith},~\bfnm{Len}\binits{L.}}
(\byear{2002}).
\btitle{Applying Lee--Carter under conditions of variable mortality decline}.
\bjournal{Popul. Stud. (Camb.)}
\bvolume{56}
\bpages{325--336}.
\bid{doi={10.1080/00324720215935}, issn={0032-4728}, pii={G6JWNNTGETJNJY2U},
  pmid={12553330}}
\bptok{imsref}%
\end{barticle}
\endbibitem

\bibitem[\protect\citeauthoryear{Booth et~al.}{2006}]{booth-etal-2006}
\begin{barticle}[author]
\bauthor{\bsnm{Booth},~\bfnm{H}\binits{H.}},
  \bauthor{\bsnm{Hyndman},~\bfnm{R~J}\binits{R.~J.}},
  \bauthor{\bsnm{Tickle},~\bfnm{L}\binits{L.}} \AND \bauthor{\bparticle{de}
  \bsnm{Jong},~\bfnm{P}\binits{P.}}
(\byear{2006}).
\btitle{Lee--Carter mortality forecasting: A multi-country comparison of
  variants and extensions}.
\bjournal{Demographic Research}
\bvolume{15}
\bpages{289--310}.
\bptok{imsref}%
\end{barticle}
\endbibitem

\bibitem[\protect\citeauthoryear{Bouchardy, Lutz and K{\"u}hni}{2011}]{krebs}
\begin{bbook}[author]
\bauthor{\bsnm{Bouchardy},~\bfnm{C}\binits{C.}},
  \bauthor{\bsnm{Lutz},~\bfnm{J-M}\binits{J.-M.}} \AND
  \bauthor{\bsnm{K{\"u}hni},~\bfnm{C}\binits{C.}}
(\byear{2011}).
\btitle{{Krebs in der Schweiz: Stand und Entwicklung von 1983 bis 2007}}.
\bpublisher{BFS, NICER, SKKR}, \baddress{Neuch{\^a}tel}.
\bptok{imsref}%
\end{bbook}
\endbibitem

\bibitem[\protect\citeauthoryear{Bray}{2002}]{bray-2002}
\begin{barticle}[mr]
\bauthor{\bsnm{Bray},~\bfnm{Isabelle}\binits{I.}}
(\byear{2002}).
\btitle{Application of {M}arkov chain {M}onte {C}arlo methods to projecting
  cancer incidence and mortality}.
\bjournal{J. Roy. Statist. Soc. Ser. C}
\bvolume{51}
\bpages{151--164}.
\bid{doi={10.1111/1467-9876.00260}, issn={0035-9254}, mr={1900163}}
\bptok{imsref}%
\end{barticle}
\endbibitem

\bibitem[\protect\citeauthoryear{Bray, Brennan and
  Boffetta}{2001}]{bray-etal-2001}
\begin{barticle}[author]
\bauthor{\bsnm{Bray},~\bfnm{I}\binits{I.}},
  \bauthor{\bsnm{Brennan},~\bfnm{P}\binits{P.}} \AND
  \bauthor{\bsnm{Boffetta},~\bfnm{P}\binits{P.}}
(\byear{2001}).
\btitle{Recent trends and future projections of lymphoid neoplasms---a
  {B}ayesian age-period-cohort analysis}.
\bjournal{Cancer Causes and Control}
\bvolume{12}
\bpages{813--820}.
\bptok{imsref}%
\end{barticle}
\endbibitem

\bibitem[\protect\citeauthoryear{Brillinger}{1986}]{brillinger-1986}
\begin{barticle}[mr]
\bauthor{\bsnm{Brillinger},~\bfnm{David~R.}\binits{D.~R.}}
(\byear{1986}).
\btitle{The natural variability of vital rates and associated statistics}.
\bjournal{Biometrics}
\bvolume{42}
\bpages{693--734}.
\bid{doi={10.2307/2530689}, issn={0006-341X}, mr={0872958}}
\bptnote{check related}%
\bptok{imsref}%
\end{barticle}
\endbibitem

\bibitem[\protect\citeauthoryear{Brouhns, Denuit and
  Vermunt}{2002}]{brouhns-etal-2002}
\begin{barticle}[mr]
\bauthor{\bsnm{Brouhns},~\bfnm{Natacha}\binits{N.}},
  \bauthor{\bsnm{Denuit},~\bfnm{Michel}\binits{M.}} \AND
  \bauthor{\bsnm{Vermunt},~\bfnm{Jeroen~K.}\binits{J.~K.}}
(\byear{2002}).
\btitle{A {P}oisson log-bilinear regression approach to the construction of
  projected lifetables}.
\bjournal{Insurance Math. Econom.}
\bvolume{31}
\bpages{373--393}.
\bid{doi={10.1016/S0167-6687(02)00185-3}, issn={0167-6687}, mr={1945540}}
\bptok{imsref}%
\end{barticle}
\endbibitem

\bibitem[\protect\citeauthoryear{Butt and Haberman}{2009}]{butt-haberman-2009}
\begin{bmisc}[author]
\bauthor{\bsnm{Butt},~\bfnm{Z}\binits{Z.}} \AND
  \bauthor{\bsnm{Haberman},~\bfnm{S}\binits{S.}}
(\byear{2009}).
\bhowpublished{ilc: A collection of {R} functions for fitting a class of
  {L}ee--{C}arter mortality models using iterative fitting algorithms.
Technical report, Actuarial Research Paper No. 190, City Univ.
London, UK.}
\bptok{imsref}%
\end{bmisc}
\endbibitem

\bibitem[\protect\citeauthoryear{Byers and Besag}{2000}]{byers-besag-2000}
\begin{barticle}[pbm]
\bauthor{\bsnm{Byers},~\bfnm{S.}\binits{S.}} \AND
  \bauthor{\bsnm{Besag},~\bfnm{J.}\binits{J.}}
(\byear{2000}).
\btitle{Inference on a collapsed margin in disease mapping}.
\bjournal{Stat. Med.}
\bvolume{19}
\bpages{2243--2249}.
\bid{issn={0277-6715},
  pii={10.1002/1097-0258(20000915/30)19:17/18<2243::AID-SIM566>3.0.CO;2-K},
  pmid={10960850}}
\bptok{imsref}%
\end{barticle}
\endbibitem

\bibitem[\protect\citeauthoryear{Carlin and
  Banerjee}{2003}]{carlin-banerjee-2003}
\begin{bincollection}[mr]
\bauthor{\bsnm{Carlin},~\bfnm{Bradley~P.}\binits{B.~P.}} \AND
  \bauthor{\bsnm{Banerjee},~\bfnm{Sudipto}\binits{S.}}
(\byear{2003}).
\btitle{Hierarchical multivariate {CAR} models for spatio-temporally correlated
  survival data}.
In \bbooktitle{Bayesian Statistics, 7 ({T}enerife, 2002)}
(\beditor{\bfnm{J~M}\binits{J.~M.}~\bsnm{Bernardo}},
 \beditor{\bfnm{M~J}\binits{M.~J.}~\bsnm{Bayarri}},
 \beditor{\bfnm{J~O}\binits{J.~O.}~\bsnm{Berger}},
 \beditor{\bfnm{A~P}\binits{A.~P.}~\bsnm{Dawid}},
 \beditor{\bfnm{D}\binits{D.}~\bsnm{Heckerman}} \AND
 \beditor{\bfnm{A~F~M}\binits{A.~F.~M.}~\bsnm{Smith}}, eds.)
\bpages{45--63}.
\bpublisher{Oxford Univ. Press}, \baddress{New York}.
\bid{mr={2003166}}
\bptnote{check related}%
\bptok{imsref}%
\end{bincollection}
\endbibitem

\bibitem[\protect\citeauthoryear{Clayton and
  Schifflers}{1987}]{clayton-schifflers-1987}
\begin{barticle}[pbm]
\bauthor{\bsnm{Clayton},~\bfnm{D.}\binits{D.}} \AND
  \bauthor{\bsnm{Schifflers},~\bfnm{E.}\binits{E.}}
(\byear{1987}).
\btitle{Models for temporal variation in cancer rates. II: Age-period-cohort
  models}.
\bjournal{Stat. Med.}
\bvolume{6}
\bpages{469--481}.
\bid{issn={0277-6715}, pmid={3629048}}
\bptok{imsref}%
\end{barticle}
\endbibitem

\bibitem[\protect\citeauthoryear{Currie, Durban and
  Eilers}{2004}]{currie-etal-2004}
\begin{barticle}[mr]
\bauthor{\bsnm{Currie},~\bfnm{Iain~D.}\binits{I.~D.}},
  \bauthor{\bsnm{Durban},~\bfnm{Maria}\binits{M.}} \AND
  \bauthor{\bsnm{Eilers},~\bfnm{Paul H.~C.}\binits{P.~H.~C.}}
(\byear{2004}).
\btitle{Smoothing and forecasting mortality rates}.
\bjournal{Stat. Model.}
\bvolume{4}
\bpages{279--298}.
\bid{doi={10.1191/1471082X04st080oa}, issn={1471-082X}, mr={2086492}}
\bptok{imsref}%
\end{barticle}
\endbibitem

\bibitem[\protect\citeauthoryear{Czado, Gneiting and
  Held}{2009}]{czado-etal-2009}
\begin{barticle}[mr]
\bauthor{\bsnm{Czado},~\bfnm{Claudia}\binits{C.}},
  \bauthor{\bsnm{Gneiting},~\bfnm{Tilmann}\binits{T.}} \AND
  \bauthor{\bsnm{Held},~\bfnm{Leonhard}\binits{L.}}
(\byear{2009}).
\btitle{Predictive model assessment for count data}.
\bjournal{Biometrics}
\bvolume{65}
\bpages{1254--1261}.
\bid{doi={10.1111/j.1541-0420.2009.01191.x}, issn={0006-341X}, mr={2756513}}
\bptok{imsref}%
\end{barticle}
\endbibitem

\bibitem[\protect\citeauthoryear{Dockery and Pope}{1994}]{dockery-pope-1994}
\begin{barticle}[pbm]
\bauthor{\bsnm{Dockery},~\bfnm{D.~W.}\binits{D.~W.}} \AND
  \bauthor{\bsnm{Pope},~\bfnm{C.~A.}\binits{C.~A.}}
(\byear{1994}).
\btitle{Acute respiratory effects of particulate air pollution}.
\bjournal{Annu. Rev. Public Health}
\bvolume{15}
\bpages{107--132}.
\bid{doi={10.1146/annurev.pu.15.050194.000543}, issn={0163-7525},
  pmid={8054077}}
\bptok{imsref}%
\end{barticle}
\endbibitem

\bibitem[\protect\citeauthoryear{Ess et~al.}{2010}]{ess-etal-2010}
\begin{barticle}[pbm]
\bauthor{\bsnm{Ess},~\bfnm{S.}\binits{S.}},
  \bauthor{\bsnm{Savidan},~\bfnm{A.}\binits{A.}},
  \bauthor{\bsnm{Frick},~\bfnm{H.}\binits{H.}},
  \bauthor{\bsnm{Rageth},~\bfnm{Ch}\binits{C.}},
  \bauthor{\bsnm{Vlastos},~\bfnm{G.}\binits{G.}},
  \bauthor{\bsnm{L{\"{u}}tolf},~\bfnm{U.}\binits{U.}} \AND
  \bauthor{\bsnm{Th{\"{u}}rlimann},~\bfnm{B.}\binits{B.}}
(\byear{2010}).
\btitle{Geographic variation in breast cancer care in Switzerland}.
\bjournal{Cancer Epidemiol.}
\bvolume{34}
\bpages{116--121}.
\bid{doi={10.1016/j.canep.2010.01.008}, issn={1877-783X},
  pii={S1877-7821(10)00018-4}, pmid={20185382}}
\bptok{imsref}%
\end{barticle}
\endbibitem

\bibitem[\protect\citeauthoryear{Fahrmeir and Tutz}{2001}]{Fahrmeir-tutz-2001}
\begin{bbook}[mr]
\bauthor{\bsnm{Fahrmeir},~\bfnm{Ludwig}\binits{L.}} \AND
  \bauthor{\bsnm{Tutz},~\bfnm{Gerhard}\binits{G.}}
(\byear{2001}).
\btitle{Multivariate Statistical Modelling Based on Generalized Linear Models},
\bedition{2nd} ed.
\bpublisher{Springer}, \baddress{New York}.
\bid{mr={1832899}}
\bptok{imsref}%
\end{bbook}
\endbibitem

\bibitem[\protect\citeauthoryear{Fienberg and
  Mason}{1979}]{fienberg-mason-1979}
\begin{barticle}[author]
\bauthor{\bsnm{Fienberg},~\bfnm{S~E}\binits{S.~E.}} \AND
  \bauthor{\bsnm{Mason},~\bfnm{W~M}\binits{W.~M.}}
(\byear{1979}).
\btitle{Identification and estimation of age-period-cohort models in the
  analysis of discrete archival data}.
\bjournal{Sociological Methodology}
\bvolume{10}
\bpages{1--67}.
\bptok{imsref}%
\end{barticle}
\endbibitem

\bibitem[\protect\citeauthoryear{Fisher}{1958}]{Fisher-1958}
\begin{bbook}[author]
\bauthor{\bsnm{Fisher},~\bfnm{R~A}\binits{R.~A.}}
(\byear{1958}).
\btitle{Statistical Methods for Research Workers},
\bedition{13th (rev.)} ed.
\bpublisher{Oliver and Boyd}, \baddress{Edinburgh}.
\bptok{imsref}%
\end{bbook}
\endbibitem

\bibitem[\protect\citeauthoryear{Fu}{2000}]{fu-2000}
\begin{barticle}[author]
\bauthor{\bsnm{Fu},~\bfnm{W~J~J}\binits{W.~J.~J.}}
(\byear{2000}).
\btitle{Ridge estimator in singular design with application to
  age-period-cohort analysis of disease rates}.
\bjournal{Comm. Statist. Theory Methods}
\bvolume{29}
\bpages{263--278}.
\bptok{imsref}%
\end{barticle}
\endbibitem

\bibitem[\protect\citeauthoryear{Gelfand and Ghosh}{1998}]{gelfand-ghosh-1998}
\begin{barticle}[mr]
\bauthor{\bsnm{Gelfand},~\bfnm{Alan~E.}\binits{A.~E.}} \AND
  \bauthor{\bsnm{Ghosh},~\bfnm{Sujit~K.}\binits{S.~K.}}
(\byear{1998}).
\btitle{Model choice: A minimum posterior predictive loss approach}.
\bjournal{Biometrika}
\bvolume{85}
\bpages{1--11}.
\bid{doi={10.1093/biomet/85.1.1}, issn={0006-3444}, mr={1627258}}
\bptok{imsref}%
\end{barticle}
\endbibitem

\bibitem[\protect\citeauthoryear{Gelfand and
  Vounatsou}{2003}]{gelfand-vounatsou-2003}
\begin{barticle}[pbm]
\bauthor{\bsnm{Gelfand},~\bfnm{Alan~E.}\binits{A.~E.}} \AND
  \bauthor{\bsnm{Vounatsou},~\bfnm{Penelope}\binits{P.}}
(\byear{2003}).
\btitle{Proper multivariate conditional autoregressive models for spatial data
  analysis}.
\bjournal{Biostatistics}
\bvolume{4}
\bpages{11--25}.
\bid{doi={10.1093/biostatistics/4.1.11}, issn={1465-4644}, pii={4/1/11},
  pmid={12925327}}
\bptok{imsref}%
\end{barticle}
\endbibitem

\bibitem[\protect\citeauthoryear{Gneiting and
  Raftery}{2007}]{gneiting-raftery-2007}
\begin{barticle}[mr]
\bauthor{\bsnm{Gneiting},~\bfnm{Tilmann}\binits{T.}} \AND
  \bauthor{\bsnm{Raftery},~\bfnm{Adrian~E.}\binits{A.~E.}}
(\byear{2007}).
\btitle{Strictly proper scoring rules, prediction, and estimation}.
\bjournal{J. Amer. Statist. Assoc.}
\bvolume{102}
\bpages{359--378}.
\bid{doi={10.1198/016214506000001437}, issn={0162-1459}, mr={2345548}}
\bptok{imsref}%
\end{barticle}
\endbibitem

\bibitem[\protect\citeauthoryear{Greco and
  Trivisano}{2009}]{greco-trivisano-2009}
\begin{barticle}[mr]
\bauthor{\bsnm{Greco},~\bfnm{Fedele~P.}\binits{F.~P.}} \AND
  \bauthor{\bsnm{Trivisano},~\bfnm{Carlo}\binits{C.}}
(\byear{2009}).
\btitle{A multivariate {CAR} model for improving the estimation of relative
  risks}.
\bjournal{Stat. Med.}
\bvolume{28}
\bpages{1707--1724}.
\bid{doi={10.1002/sim.3577}, issn={0277-6715}, mr={2675246}}
\bptok{imsref}%
\end{barticle}
\endbibitem

\bibitem[\protect\citeauthoryear{Hansell}{2004}]{hansell-phd-2004}
\begin{bmisc}[author]
\bauthor{\bsnm{Hansell},~\bfnm{A~L}\binits{A.~L.}}
(\byear{2004}).
\bhowpublished{The {e}pidemiology of {c}hronic {o}bstructive {p}ulmonary {d}isease in
the {UK}: {s}patial and {t}emporal {v}ariations.
Ph.D. thesis, Faculty of Medicine,  Univ. London,
Imperial College, St Mary's Campus}.
\bptok{imsref}%
\end{bmisc}
\endbibitem

\bibitem[\protect\citeauthoryear{Hansell et~al.}{2003}]{hansell-etal-2003}
\begin{barticle}[author]
\bauthor{\bsnm{Hansell},~\bfnm{A}\binits{A.}},
  \bauthor{\bsnm{Knorr-Held},~\bfnm{L}\binits{L.}},
  \bauthor{\bsnm{Best},~\bfnm{N}\binits{N.}},
  \bauthor{\bsnm{Schmid},~\bfnm{V}\binits{V.}} \AND
  \bauthor{\bsnm{Aylin},~\bfnm{P}\binits{P.}}
(\byear{2003}).
\btitle{{COPD} mortality trends 1950--1999 in {E}ngland and
Wales---Did the 1956
  {C}lean {A}ir {A}ct make a detectable difference?}
\bjournal{Epidemiology}
\bvolume{14}
\bpages{S55}.
\bptok{imsref}%
\end{barticle}
\endbibitem

\bibitem[\protect\citeauthoryear{Harvey}{1990}]{Harvey-1990}
\begin{bbook}[author]
\bauthor{\bsnm{Harvey},~\bfnm{{Andrew C. }}\binits{A.}}
(\byear{1990}).
\btitle{Forecasting, Structural Time Series Models and the Kalman Filter},
  \bedition{Reprinted} ed.
\bpublisher{Cambridge Univ. Press}, \baddress{Cambridge}.
\bptok{imsref}%
\end{bbook}
\endbibitem

\bibitem[\protect\citeauthoryear{Held and Riebler}{2011}]{held-riebler-2010}
\begin{bmisc}[pbm]
\bauthor{\bsnm{Held},~\bfnm{Leonhard}\binits{L.}} \AND
  \bauthor{\bsnm{Riebler},~\bfnm{Andrea}\binits{A.}}
(\byear{2011}).
\bhowpublished{A conditional approach for inference in multivariate age-period-cohort
  models.
\textit{Stat. Methods Med. Res.} To appear. DOI:
\href{http://dx.doi.org/10.1177/0962280210379761}{10.1177/0962280210379761}.}
\bid{doi={10.1177/0962280210379761}, issn={1477-0334}, pii={0962280210379761},
  pmid={20826502}}
\bptok{imsref}%
\end{bmisc}
\endbibitem

\bibitem[\protect\citeauthoryear{Heuer}{1997}]{heuer-1997}
\begin{barticle}[pbm]
\bauthor{\bsnm{Heuer},~\bfnm{C.}\binits{C.}}
(\byear{1997}).
\btitle{Modeling of time trends and interactions in vital rates using
  restricted regression splines}.
\bjournal{Biometrics}
\bvolume{53}
\bpages{161--177}.
\bid{issn={0006-341X}, pmid={9147590}}
\bptok{imsref}%
\end{barticle}
\endbibitem

\bibitem[\protect\citeauthoryear{Holford}{1983}]{holford-1983}
\begin{barticle}[mr]
\bauthor{\bsnm{Holford},~\bfnm{Theodore~R.}\binits{T.~R.}}
(\byear{1983}).
\btitle{The estimation of age, period and cohort effects for vital rates}.
\bjournal{Biometrics}
\bvolume{39}
\bpages{311--324}.
\bid{doi={10.2307/2531004}, issn={0006-341X}, mr={0714415}}
\bptok{imsref}%
\end{barticle}
\endbibitem

\bibitem[\protect\citeauthoryear{Holford}{1992}]{holford-1992}
\begin{barticle}[pbm]
\bauthor{\bsnm{Holford},~\bfnm{T.~R.}\binits{T.~R.}}
(\byear{1992}).
\btitle{Analysing the temporal effects of age, period and cohort}.
\bjournal{Stat. Methods Med. Res.}
\bvolume{1}
\bpages{317--337}.
\bid{issn={0962-2802}, pmid={1341663}}
\bptok{imsref}%
\end{barticle}
\endbibitem

\bibitem[\protect\citeauthoryear{Holford}{2006}]{holford-2006}
\begin{barticle}[mr]
\bauthor{\bsnm{Holford},~\bfnm{Theodore~R.}\binits{T.~R.}}
(\byear{2006}).
\btitle{Approaches to fitting age-period-cohort models with unequal intervals}.
\bjournal{Stat. Med.}
\bvolume{25}
\bpages{977--993}.
\bid{doi={10.1002/sim.2253}, issn={0277-6715}, mr={2225187}}
\bptok{imsref}%
\end{barticle}
\endbibitem

\bibitem[\protect\citeauthoryear{Human Mortality Database}{2011}]{hmd-2010}
\begin{bmisc}[author]
\borganization{Human Mortality Database}
(\byear{2011}).
\bhowpublished{Univ. California, Berkeley (USA),
and Max Planck Institute for Demographic Research (Germany).
Available at \href{http://www.mortality.org}{www.mortality.org}
or \href{http://www.humanmortality.de}{www.humanmortality.de}.}
\bptok{imsref}%
\end{bmisc}
\endbibitem

\bibitem[\protect\citeauthoryear{Jacobsen et~al.}{2004}]{jacobsen-etal-2004}
\begin{barticle}[author]
\bauthor{\bsnm{Jacobsen},~\bfnm{R}\binits{R.}},
  \bauthor{\bsnm{Von~Euler},~\bfnm{M}\binits{M.}},
  \bauthor{\bsnm{Osler},~\bfnm{M}\binits{M.}},
  \bauthor{\bsnm{Lynge},~\bfnm{E}\binits{E.}} \AND
  \bauthor{\bsnm{Keiding},~\bfnm{N}\binits{N.}}
(\byear{2004}).
\btitle{Women's death in {S}candinavia---what makes {D}enmark different?}
\bjournal{European Journal of Epidemiology}
\bvolume{19}
\bpages{117--121}.
\bptok{imsref}%
\end{barticle}
\endbibitem

\bibitem[\protect\citeauthoryear{Kazerouni et~al.}{2004}]{kazerouni-etal-2004}
\begin{barticle}[author]
\bauthor{\bsnm{Kazerouni},~\bfnm{N.}\binits{N.}},
  \bauthor{\bsnm{Alverson},~\bfnm{C.~J.}\binits{C.~J.}},
  \bauthor{\bsnm{Redd},~\bfnm{S.~C.}\binits{S.~C.}},
  \bauthor{\bsnm{Mott},~\bfnm{J.~A.}\binits{J.~A.}} \AND
  \bauthor{\bsnm{Mannino},~\bfnm{D.~M.}\binits{D.~M.}}
(\byear{2004}).
\btitle{Sex differences in {COPD} and lung cancer mortality
trends---{U}nited
  {S}tates, 1968--1999}.
\bjournal{J. Women's Health}
\bvolume{13}
\bpages{17--23}.
\bptok{imsref}%
\end{barticle}
\endbibitem

\bibitem[\protect\citeauthoryear{Knorr-Held}{2000}]{knorrHeld-2000}
\begin{barticle}[pbm]
\bauthor{\bsnm{Knorr-Held},~\bfnm{L.}\binits{L.}}
(\byear{2000}).
\btitle{Bayesian modelling of inseparable space--time variation in disease
  risk}.
\bjournal{Stat. Med.}
\bvolume{19}
\bpages{2555--2567}.
\bptok{imsref}%
\end{barticle}
\endbibitem

\bibitem[\protect\citeauthoryear{Knorr-Held and
  Rainer}{2001}]{knorrHeld-rainer-2001}
\begin{barticle}[pbm]
\bauthor{\bsnm{Knorr-Held},~\bfnm{L.}\binits{L.}} \AND
  \bauthor{\bsnm{Rainer},~\bfnm{E.}\binits{E.}}
(\byear{2001}).
\btitle{Projections of lung cancer mortality in West Germany: A case study in
  Bayesian prediction}.
\bjournal{Biostatistics}
\bvolume{2}
\bpages{109--129}.
\bid{doi={10.1093/biostatistics/2.1.109}, issn={1468-4357}, pii={2/1/109},
  pmid={12933560}}
\bptok{imsref}%
\end{barticle}
\endbibitem

\bibitem[\protect\citeauthoryear{Kolte et~al.}{2008}]{kolte-etal-2008}
\begin{barticle}[pbm]
\bauthor{\bsnm{Kolte},~\bfnm{Ida~Viktoria}\binits{I.~V.}},
  \bauthor{\bsnm{Skinh{\o}j},~\bfnm{Peter}\binits{P.}},
  \bauthor{\bsnm{Keiding},~\bfnm{Niels}\binits{N.}} \AND
  \bauthor{\bsnm{Lynge},~\bfnm{Elsebeth}\binits{E.}}
(\byear{2008}).
\btitle{The Spanish flu in Denmark}.
\bjournal{Scand. J. Infect. Dis.}
\bvolume{40}
\bpages{538--546}.
\bid{doi={10.1080/00365540701870903}, issn={0036-5548}, pii={790201977},
  pmid={18584544}}
\bptok{imsref}%
\end{barticle}
\endbibitem

\bibitem[\protect\citeauthoryear{Konishi}{1985}]{konishi-1985}
\begin{barticle}[mr]
\bauthor{\bsnm{Konishi},~\bfnm{Sadanori}\binits{S.}}
(\byear{1985}).
\btitle{Normalizing and variance stabilizing transformations for intraclass
  correlations}.
\bjournal{Ann. Inst. Statist. Math.}
\bvolume{37}
\bpages{87--94}.
\bid{doi={10.1007/BF02481082}, issn={0020-3157}, mr={0790377}}
\bptok{imsref}%
\end{barticle}
\endbibitem

\bibitem[\protect\citeauthoryear{Kuang, Nielsen and
  Nielsen}{2008}]{kuang-etal-2008}
\begin{barticle}[mr]
\bauthor{\bsnm{Kuang},~\bfnm{D.}\binits{D.}},
  \bauthor{\bsnm{Nielsen},~\bfnm{B.}\binits{B.}} \AND
  \bauthor{\bsnm{Nielsen},~\bfnm{J.~P.}\binits{J.~P.}}
(\byear{2008}).
\btitle{Identification of the age-period-cohort model and the extended
  chain-ladder model}.
\bjournal{Biometrika}
\bvolume{95}
\bpages{979--986}.
\bid{doi={10.1093/biomet/asn026}, issn={0006-3444}, mr={2461224}}
\bptok{imsref}%
\end{barticle}
\endbibitem

\bibitem[\protect\citeauthoryear{Lagazio, Biggeri and
  Dreassi}{2003}]{lagazio-etal-2003}
\begin{barticle}[author]
\bauthor{\bsnm{Lagazio},~\bfnm{C.}\binits{C.}},
  \bauthor{\bsnm{Biggeri},~\bfnm{A.}\binits{A.}} \AND
  \bauthor{\bsnm{Dreassi},~\bfnm{E.}\binits{E.}}
(\byear{2003}).
\btitle{Age-period-cohort models and disease mapping}.
\bjournal{Environmetrics}
\bvolume{14}
\bpages{475--490}.
\bptok{imsref}%
\end{barticle}
\endbibitem

\bibitem[\protect\citeauthoryear{Lee and Carter}{1992}]{lee-carter-1992}
\begin{barticle}[author]
\bauthor{\bsnm{Lee},~\bfnm{R.~D.}\binits{R.~D.}} \AND
  \bauthor{\bsnm{Carter},~\bfnm{L.~R.}\binits{L.~R.}}
(\byear{1992}).
\btitle{Modeling and forecasting {U.S.} mortality}.
\bjournal{J. Amer. Statist. Assoc.}
\bvolume{87}
\bpages{659--671}.
\bptok{imsref}%
\end{barticle}
\endbibitem

\bibitem[\protect\citeauthoryear{Lehmann}{1999}]{Lehmann-1999}
\begin{bbook}[mr]
\bauthor{\bsnm{Lehmann},~\bfnm{E.~L.}\binits{E.~L.}}
(\byear{1999}).
\btitle{Elements of Large-Sample Theory}.
\bpublisher{Springer}, \baddress{New York}.
\bid{doi={10.1007/b98855}, mr={1663158}}
\bptok{imsref}%
\end{bbook}
\endbibitem

\bibitem[\protect\citeauthoryear{Levi et~al.}{1993}]{levi-etal-1993}
\begin{barticle}[pbm]
\bauthor{\bsnm{Levi},~\bfnm{F.}\binits{F.}},
  \bauthor{\bsnm{Randimbison},~\bfnm{L.}\binits{L.}},
  \bauthor{\bsnm{Te},~\bfnm{V.~C.}\binits{V.~C.}},
  \bauthor{\bsnm{Rolland-Portal},~\bfnm{I.}\binits{I.}},
  \bauthor{\bsnm{Franceschi},~\bfnm{S.}\binits{S.}} \AND
  \bauthor{\bsnm{La~Vecchia},~\bfnm{C.}\binits{C.}}
(\byear{1993}).
\btitle{Multiple primary cancers in the Vaud cancer registry, Switzerland,
  1974-89}.
\bjournal{Br. J. Cancer}
\bvolume{67}
\bpages{391--395}.
\bid{issn={0007-0920}, pmcid={1968177}, pmid={8431373}}
\bptok{imsref}%
\end{barticle}
\endbibitem

\bibitem[\protect\citeauthoryear{Levi et~al.}{1998}]{levi-etal-1998}
\begin{barticle}[pbm]
\bauthor{\bsnm{Levi},~\bfnm{F.}\binits{F.}},
  \bauthor{\bsnm{La~Vecchia},~\bfnm{C.}\binits{C.}},
  \bauthor{\bsnm{Randimbison},~\bfnm{L.}\binits{L.}},
  \bauthor{\bsnm{Erler},~\bfnm{G.}\binits{G.}},
  \bauthor{\bsnm{Te},~\bfnm{V.~C.}\binits{V.~C.}} \AND
  \bauthor{\bsnm{Franceschi},~\bfnm{S.}\binits{S.}}
(\byear{1998}).
\btitle{Incidence, mortality and survival from prostate cancer in Vaud and
  Neuch\^{a}tel, Switzerland, 1974--1994}.
\bjournal{Ann. Oncol.}
\bvolume{9}
\bpages{31--35}.
\bid{issn={0923-7534}, pmid={9541680}}
\bptok{imsref}%
\end{barticle}
\endbibitem

\bibitem[\protect\citeauthoryear{Levi et~al.}{2002}]{levi-etal-2002}
\begin{barticle}[pbm]
\bauthor{\bsnm{Levi},~\bfnm{Fabio}\binits{F.}},
  \bauthor{\bsnm{Randimbison},~\bfnm{Lalao}\binits{L.}},
  \bauthor{\bsnm{Te},~\bfnm{Van-Cong}\binits{V.-C.}} \AND
  \bauthor{\bsnm{La~Vecchia},~\bfnm{Carlo}\binits{C.}}
(\byear{2002}).
\btitle{Thyroid cancer in Vaud, Switzerland: An update}.
\bjournal{Thyroid}
\bvolume{12}
\bpages{163--168}.
\bid{doi={10.1089/105072502753522400}, issn={1050-7256}, pmid={11916286}}
\bptok{imsref}%
\end{barticle}
\endbibitem

\bibitem[\protect\citeauthoryear{Li and Lee}{2005}]{li-lee-2005}
\begin{barticle}[author]
\bauthor{\bsnm{Li},~\bfnm{N}\binits{N.}} \AND
  \bauthor{\bsnm{Lee},~\bfnm{R}\binits{R.}}
(\byear{2005}).
\btitle{Coherent mortality forecasts for a group of populations: An extension
  of the Lee--Carter method}.
\bjournal{Demography}
\bvolume{42}
\bpages{575--594}.
\bptok{imsref}%
\end{barticle}
\endbibitem

\bibitem[\protect\citeauthoryear{Lindley}{1965}]{Lindley-1965}
\begin{bbook}[author]
\bauthor{\bsnm{Lindley},~\bfnm{D}\binits{D.}}
(\byear{1965}).
\btitle{Introduction to Probability and Statistics from a Bayesian Viewpoint, Part 2,
Inference}.
\bpublisher{Cambridge Univ. Press}, \baddress{Cambridge}.
\bptok{imsref}%
\end{bbook}
\endbibitem

\bibitem[\protect\citeauthoryear{Mardia}{1988}]{mardia-1988}
\begin{barticle}[author]
\bauthor{\bsnm{Mardia},~\bfnm{K}\binits{K.}}
(\byear{1988}).
\btitle{Multi-dimensional multivariate {G}aussian {M}arkov random fields with
  application to image processing}.
\bjournal{J. Multivariate Anal.}
\bvolume{24}
\bpages{265--284}.
\bptok{imsref}%
\end{barticle}
\endbibitem

\bibitem[\protect\citeauthoryear{Nakamura}{1986}]{nakamura-1986}
\begin{barticle}[author]
\bauthor{\bsnm{Nakamura},~\bfnm{T}\binits{T.}}
(\byear{1986}).
\btitle{Bayesian cohort models for general cohort table analyses}.
\bjournal{Ann. Inst. Statist. Math.}
\bvolume{38}
\bpages{353--370}.
\bptok{imsref}%
\end{barticle}
\endbibitem

\bibitem[\protect\citeauthoryear{Ogata et~al.}{2000}]{ogata-etal-2000}
\begin{barticle}[author]
\bauthor{\bsnm{Ogata},~\bfnm{Y}\binits{Y.}},
  \bauthor{\bsnm{Katsura},~\bfnm{K}\binits{K.}},
  \bauthor{\bsnm{Keiding},~\bfnm{N}\binits{N.}},
  \bauthor{\bsnm{Holst},~\bfnm{C}\binits{C.}} \AND
  \bauthor{\bsnm{Green},~\bfnm{A}\binits{A.}}
(\byear{2000}).
\btitle{Empirical {B}ayes age-period-cohort analysis of retrospective incidence
  data}.
\bjournal{Scand. J. Stat.}
\bvolume{27}
\bpages{415--432}.
\bptok{imsref}%
\end{barticle}
\endbibitem

\bibitem[\protect\citeauthoryear{Osmond and
  Gardner}{1982}]{osmond-gardner-1982}
\begin{barticle}[pbm]
\bauthor{\bsnm{Osmond},~\bfnm{C.}\binits{C.}} \AND
  \bauthor{\bsnm{Gardner},~\bfnm{M.~J.}\binits{M.~J.}}
(\byear{1982}).
\btitle{Age, period and cohort models applied to cancer mortality rates}.
\bjournal{Stat. Med.}
\bvolume{1}
\bpages{245--259}.
\bid{issn={0277-6715}, pmid={7187097}}
\bptok{imsref}%
\end{barticle}
\endbibitem

\bibitem[\protect\citeauthoryear{Paul et~al.}{2010}]{paul-etal-2010}
\begin{barticle}[mr]
\bauthor{\bsnm{Paul},~\bfnm{M.}\binits{M.}},
  \bauthor{\bsnm{Riebler},~\bfnm{A.}\binits{A.}},
  \bauthor{\bsnm{Bachmann},~\bfnm{L.~M.}\binits{L.~M.}},
  \bauthor{\bsnm{Rue},~\bfnm{H.}\binits{H.}} \AND
  \bauthor{\bsnm{Held},~\bfnm{L.}\binits{L.}}
(\byear{2010}).
\btitle{Bayesian bivariate meta-analysis of diagnostic test studies using
  integrated nested {L}aplace approximations}.
\bjournal{Stat. Med.}
\bvolume{29}
\bpages{1325--1339}.
\bid{doi={10.1002/sim.3858}, issn={0277-6715}, mr={2757228}}
\bptok{imsref}%
\end{barticle}
\endbibitem

\bibitem[\protect\citeauthoryear{R Development Core Team}{2010}]{r-project}
\begin{bmisc}[author]
\borganization{R Development Core Team}
(\byear{2010}).
\bhowpublished{R: A Language and Environment for Statistical Computing. R
  Foundation for Statistical Computing, Vienna, Austria}.
\bptok{imsref}%
\end{bmisc}
\endbibitem

\bibitem[\protect\citeauthoryear{Riebler and Held}{2010}]{riebler-held-2010}
\begin{barticle}[author]
\bauthor{\bsnm{Riebler},~\bfnm{A}\binits{A.}} \AND
  \bauthor{\bsnm{Held},~\bfnm{L}\binits{L.}}
(\byear{2010}).
\btitle{The analysis of heterogeneous time trends in multivariate
  age-period-cohort models}.
\bjournal{Biostatistics}
\bvolume{11}
\bpages{57--69}.
\bptok{imsref}%
\end{barticle}
\endbibitem

\bibitem[\protect\citeauthoryear{Riebler, Held and
  Rue}{2011}]{riebler-etal-2011supp}
\begin{bmisc}[author]
\bauthor{\bsnm{Riebler},~\bfnm{A}\binits{A.}},
  \bauthor{\bsnm{Held},~\bfnm{L}\binits{L.}} \AND
  \bauthor{\bsnm{Rue},~\bfnm{H}\binits{H.}}
(\byear{2011}).
\bhowpublished{Supplement to ``Estimation and extrapolation of time trends in registry
data---Borrowing strength from related populations.''
\href{http://dx.doi.org/10.1214/11-AOAS498SUPP}{DOI:10.1214/11-AOAS498SUPP}}.
\bptok{imsref}%
\end{bmisc}
\endbibitem

\bibitem[\protect\citeauthoryear{Riebler et~al.}{2011}]{riebler-etal-2011}
\begin{bmisc}[author]
\bauthor{\bsnm{Riebler},~\bfnm{A}\binits{A.}},
  \bauthor{\bsnm{Held},~\bfnm{L}\binits{L.}},
  \bauthor{\bsnm{Rue},~\bfnm{H}\binits{H.}} \AND
  \bauthor{\bsnm{Bopp},~\bfnm{M}\binits{M.}}
(\byear{2011}).
\bhowpublished{Gender-specific differences and the impact of family integration on
  time trends in age-stratified {S}wiss suicide rates.
\textit{J.~Roy. Statist. Soc. Ser.~A}.
To appear.}
\bptok{imsref}%
\end{bmisc}
\endbibitem

\bibitem[\protect\citeauthoryear{Robertson and
  Boyle}{1986}]{robertson-boyle-1986}
\begin{barticle}[pbm]
\bauthor{\bsnm{Robertson},~\bfnm{C.}\binits{C.}} \AND
  \bauthor{\bsnm{Boyle},~\bfnm{P.}\binits{P.}}
(\byear{1986}).
\btitle{Age, period and cohort models: The use of individual records}.
\bjournal{Stat. Med.}
\bvolume{5}
\bpages{527--538}.
\bid{issn={0277-6715}, pmid={3675723}}
\bptok{imsref}%
\end{barticle}
\endbibitem

\bibitem[\protect\citeauthoryear{Rue and Held}{2005}]{GMRFbook}
\begin{bbook}[mr]
\bauthor{\bsnm{Rue},~\bfnm{H{\aa}vard}\binits{H.}} \AND
  \bauthor{\bsnm{Held},~\bfnm{Leonhard}\binits{L.}}
(\byear{2005}).
\btitle{Gaussian {M}arkov Random Fields: Theory and Applications}.
\bseries{Monographs on Statistics and Applied Probability}
\bvolume{104}.
\bpublisher{Chapman and Hall/CRC}, \baddress{Boca Raton, FL}.
\bid{doi={10.1201/9780203492024}, mr={2130347}}
\bptok{imsref}%
\end{bbook}
\endbibitem

\bibitem[\protect\citeauthoryear{Rue, Martino and Chopin}{2009}]{rue-etal-2009}
\begin{barticle}[mr]
\bauthor{\bsnm{Rue},~\bfnm{H{\aa}vard}\binits{H.}},
  \bauthor{\bsnm{Martino},~\bfnm{Sara}\binits{S.}} \AND
  \bauthor{\bsnm{Chopin},~\bfnm{Nicolas}\binits{N.}}
(\byear{2009}).
\btitle{Approximate {B}ayesian inference for latent {G}aussian models by using
  integrated nested {L}aplace approximations}.
\bjournal{J. R. Stat. Soc. Ser. B Stat. Methodol.}
\bvolume{71}
\bpages{319--392}.
\bid{doi={10.1111/j.1467-9868.2008.00700.x}, issn={1369-7412}, mr={2649602}}
\bptnote{check related}%
\bptok{imsref}%
\end{barticle}
\endbibitem

\bibitem[\protect\citeauthoryear{Schmid and Held}{2004}]{schmid-held-2004}
\begin{barticle}[mr]
\bauthor{\bsnm{Schmid},~\bfnm{Volker}\binits{V.}} \AND
  \bauthor{\bsnm{Held},~\bfnm{Leonhard}\binits{L.}}
(\byear{2004}).
\btitle{Bayesian extrapolation of space-time trends in cancer registry data}.
\bjournal{Biometrics}
\bvolume{60}
\bpages{1034--1042}.
\bid{doi={10.1111/j.0006-341X.2004.00259.x}, issn={0006-341X}, mr={2133556}}
\bptok{imsref}%
\end{barticle}
\endbibitem

\bibitem[\protect\citeauthoryear{Schmid and Held}{2007}]{schmid-held-2007}
\begin{barticle}[author]
\bauthor{\bsnm{Schmid},~\bfnm{V~J}\binits{V.~J.}} \AND
  \bauthor{\bsnm{Held},~\bfnm{L}\binits{L.}}
(\byear{2007}).
\btitle{Bayesian age-period-cohort modeling and prediction---{BAMP}}.
\bjournal{Journal of Statistical Software}
\bvolume{21}
\bpages{1--15}.
\bptok{imsref}%
\end{barticle}
\endbibitem

\bibitem[\protect\citeauthoryear{Schr{\"o}dle
  et~al.}{2011}]{schroedle-etal-2010}
\begin{barticle}[author]
\bauthor{\bsnm{Schr{\"o}dle},~\bfnm{B}\binits{B.}},
  \bauthor{\bsnm{Held},~\bfnm{L}\binits{L.}},
  \bauthor{\bsnm{Riebler},~\bfnm{A}\binits{A.}} \AND
  \bauthor{\bsnm{Danuser},~\bfnm{J}\binits{J.}}
(\byear{2011}).
\btitle{Using integrated nested Laplace approximations
for the evaluation of veterinary surveillance data from
  {S}witzerland: A case study}.
\bjournal{J. R. Stat. Soc. Ser. C. Appl. Stat.}
\bvolume{60}
\bpages{261--279}.
\bptok{imsref}%
\end{barticle}
\endbibitem

\bibitem[\protect\citeauthoryear{Sunyer}{2001}]{sunyer-2001}
\begin{barticle}[pbm]
\bauthor{\bsnm{Sunyer},~\bfnm{J.}\binits{J.}}
(\byear{2001}).
\btitle{Urban air pollution and chronic obstructive pulmonary disease: A~review}.
\bjournal{Eur. Respir. J.}
\bvolume{17}
\bpages{1024--1033}.
\bid{issn={0903-1936}, pmid={11488305}}
\bptok{imsref}%
\end{barticle}
\endbibitem

\bibitem[\protect\citeauthoryear{Verkooijhen
  et~al.}{2003}]{verkooijhen-etal-2003}
\begin{barticle}[author]
\bauthor{\bsnm{Verkooijhen},~\bfnm{H~M}\binits{H.~M.}},
  \bauthor{\bsnm{Fioretta},~\bfnm{G}\binits{G.}},
  \bauthor{\bsnm{Vlastos},~\bfnm{G}\binits{G.}},
  \bauthor{\bsnm{Morabia},~\bfnm{A}\binits{A.}},
  \bauthor{\bsnm{Schubert},~\bfnm{H}\binits{H.}},
  \bauthor{\bsnm{Sappino},~\bfnm{AP}\binits{A.}},
  \bauthor{\bsnm{Pelte},~\bfnm{MF}\binits{M.}},
  \bauthor{\bsnm{Schafer},~\bfnm{P}\binits{P.}},
  \bauthor{\bsnm{Kurtz},~\bfnm{J}\binits{J.}} \AND
  \bauthor{\bsnm{Bouchardy},~\bfnm{C}\binits{C.}}
(\byear{2003}).
\btitle{{Important increase of invasive lobular breast cancer incidence in
  Geneva, Switzerland}}.
\bjournal{International Journal of Cancer}
\bvolume{104}
\bpages{778--781}.
\bptok{imsref}%
\end{barticle}
\endbibitem

\bibitem[\protect\citeauthoryear{Yang, Fu and Land}{2004}]{yang-etal-2004}
\begin{barticle}[author]
\bauthor{\bsnm{Yang},~\bfnm{Yang}\binits{Y.}},
  \bauthor{\bsnm{Fu},~\bfnm{Wenjiang~J.}\binits{W.~J.}} \AND
  \bauthor{\bsnm{Land},~\bfnm{Kenneth~C.}\binits{K.~C.}}
(\byear{2004}).
\btitle{A methodological comparison of age-period-cohort models: {T}he
  intrinsic estimator and conventional generalized linear models}.
\bjournal{Sociological Methodology}
\bvolume{34}
\bpages{75--110}.
\bptok{imsref}%
\end{barticle}
\endbibitem

\end{thebibliography}
\end{document}